\documentclass[]{jfm} 

\usepackage{natbib}
\usepackage{epstopdf}
\usepackage{upmath}
\usepackage{amssymb}
\usepackage{amsbsy}
\usepackage{amsmath}

\usepackage[dvips]{graphicx,color}
\usepackage{textcomp}
\usepackage{rotating}
\definecolor{white}{gray}{1.00}
\definecolor{gray}{gray}{0.85}
\providecommand\bnabla{\boldsymbol{\nabla}}
\providecommand\bcdot{\boldsymbol{\cdot}}
\newcommand\biS{\boldsymbol{S}}
\newcommand\etb{\boldsymbol{\eta}}
\newcommand\beq{\begin{equation}}
\newcommand\eeq{\end{equation}}
\newcommand\Rey{\mbox{\textit{Re}}}  
\newcommand\Pran{\mbox{\textit{Pr}}} 
\newcommand\Pen{\mbox{\textit{Pe}}}  
\newcommand{\partiel}[2]{\frac{\partial #1}{\partial #2}}
\newcommand{\ppartiel}[2]{\frac{\partial^2 #1}{\partial #2^2}}
\newcommand{\ud}{\;d}
\newcommand{\hard}[2]{\frac{\text{d} #1}{\text{d} #2}}
\newcommand\slsQ{\mathsfbi{Q}} 

\newcommand\hatR{\skew3\hat{R}}      

\newsavebox{\astrutbox}
\sbox{\astrutbox}{\rule[-5pt]{0pt}{20pt}}
\newcommand{\astrut}{\usebox{\astrutbox}}

\newcommand\GaPQ{\ensuremath{G_a(P,Q)}}
\newcommand\GsPQ{\ensuremath{G_s(P,Q)}}
\newcommand\p{\ensuremath{\partial}}
\newcommand\tti{\ensuremath{\rightarrow\infty}}
\newcommand\kgd{\ensuremath{k\gamma d}}
\newcommand\shalf{\ensuremath{{\scriptstyle\frac{1}{2}}}}

\newcommand\squart{\ensuremath{{\textstyle\frac{1}{4}}}}
\newcommand\thalf{\ensuremath{{\textstyle\frac{1}{2}}}}
\newcommand\Gat{\ensuremath{\widetilde{G_a}}}
\newcommand\ttz{\ensuremath{\rightarrow 0}}
\newcommand\ndq{\ensuremath{\frac{\mbox{$\partial$}}{\mbox{$\partial$} n_q}}}
\newcommand\sumjm{\ensuremath{\sum_{j=1}^{M}}}
\newcommand\pvi{\ensuremath{\int_0^{\infty}%
  \mskip \ifCUPmtlplainloaded -30mu\else -33mu\fi -\quad}}
\newcommand\etal{\mbox{\textit{et al. }}}
\newcommand\etc{etc.\ }

\title{Osmotically driven pipe flows and their relation to sugar transport in plants}

\author[K. H. Jensen, E. Rio, R. Hansen, C. Clanet and T. Bohr]{K\AA RE H.  JENSEN$^{1,2}$, EMMANUELLE RIO $^{1,3}$
, RASMUS HANSEN$^1$, CHRISTOPHE  CLANET$^4$ \and TOMAS  BOHR$^1$}
\affiliation{
$^1$Center for Fluid Dynamics, Department of Physics, \\Technical University of Denmark, Building 309, 2800 Kgs. Lyngby, Denmark\\[\affilskip]
$^2$Center for Fluid Dynamics, Department of Micro- and Nanotechnology, \\ Technical University of Denmark, DTU Nanotech Building 345 East, 2800 Kgs. Lyngby, Denmark\\[\affilskip]
$^3$ Present address: Laboratoire de Physique des Solides, Univ. Paris-Sud, CNRS, UMR 8502, F-91405 Orsay Cedex, France.\\[\affilskip]
$^4$IRPHE, UniversitŽs d'Aix-Marseille, 49 Rue Fr\'ed\'eric Joliot-Curie \\BP 146, F-13384 Marseille cedex 13, France}
\begin{document}
\maketitle
\begin{abstract}
In plants, osmotically driven flows are believed to be responsible for translocation of sugar in the pipe-like phloem cell network, spanning the entire length of the plant. In this paper, we present an experimental and theoretical study of transient osmotically driven flows through pipes with semipermeable walls. We extend the experimental work of Eschrich, Evert and Young \cite[]{Eschrich:1972} by providing a more accurate version of their experiment allowing for better comparison with theory.
In the experiments we measure the dynamics and structure of a ``sugar front", i.e. the transport and decay of a sudden loading of sugar in a pipe which is closed in both ends. We include measurements of pressure inside the membrane tube allowing us to compare the experiments directly with theory
and, in particular, to confirm quantitatively the exponential decay of the front in a closed tube.
In a novel setup we are able to measure the entire concentration profile as the sugar front moves. In contrast to previous studies we find very good agreement between experiment and theory.

In the limit of low axial resistance (valid in our experiments as well as in many cases in plants) we show that the equations can be solved exactly by the method of characteristics yielding, in general, an implicit solution.
Further we show that under more general conditions the equations of motion can be rewritten as a single integro-differential equation, which can be readily solved numerically.  The applicability of our results to plants is discussed and it is shown that it is probable that the pressure-flow hypothesis can account for short distance transport of sugar in plants.
\end{abstract}

\section{Introduction}
\begin{figure}
\centerline{\includegraphics[width=1\textwidth]{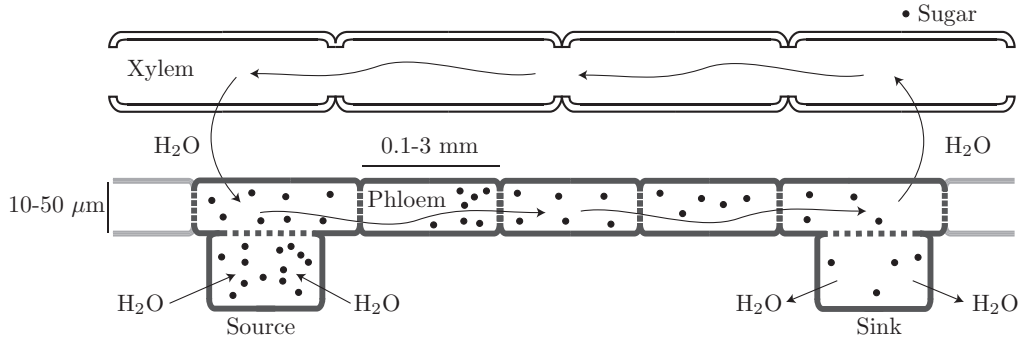}}
\caption{In plants, two separate pipe-like systems are responsible for the transport of water and sugar. The xylem conducts water from the roots to the shoot while the phloem conducts sugar and other nutrients from places of production to places of growth and storage.
The mechanism believed to be responsible for sugar-translocation in the phloem, called the M\" unch mechanism or the pressure-flow hypothesis \cite[]{Nobel:1999}, states that following: As sugar
is produced via photosynthesis in sources it is actively loaded into the tubular phloem cells. As it enters the phloem, the chemical potential of the water inside is lowered compared to the surrounding tissue, thereby creating a net flux of water into the phloem cells. This in-flux of water in turn creates a bulk flow of sugar and water towards the sugar sink shown on the right, where active unloading takes place. As the sugar is removed, the chemical potential of the water inside the phloem is raised resulting in a flow of water out of the sieve element.}
\label{fig:plant}
\end{figure}

The translocation of sugar in plants, which takes place in the phloem sieve tubes, is not well understood on the quantitative level. The current belief, called the pressure flow hypothesis \cite[]{Nobel:1999}, is based on the pioneering work of Ernst M\"unch in the 1920'ies \cite[see eg.][]{Munch:1930}. It states, that the motion in the phloem is purely passive, due to the osmotic pressures that build up relatively to the neighboring xylem as a response of loading and unloading of sugar in different parts of the plant, as shown in figure \ref{fig:plant}. This mechanism is much more effective than diffusion, since the osmotic pressure differences caused by different sugar concentrations in the phloem create a bulk flow directed from large concentrations to small concentrations, in accordance with the basic need of the plant.
Such flows are often called Osmotically Driven Pressure Flows \cite[]{Thompson:2003}, or Osmotically Driven Volume Flows \cite[]{Eschrich:1972}.

It is, however, not clear how well this mechanism is able to account for the sugar translocation in plants on the quantitative level . Since the sieve tube elements that make up the phloem are living cells, the picture can indeed be much more complicated. For a large tree it would thus seem improbable that sugar transport e.g. from leaf to root by this mechanism would be sufficiently efficient, and in this case active transport processes might play an important role. On the other hand, transport over short distances, e.g. locally in leaves or from a leaf to a nearby shoot might be more convincingly described by the pressure-flow hypothesis. In any case, we need a better understanding of such flows in order to decide, whether they compare sensibly with translocation in plants, and this is the aim of the present paper. In particular, we have chosen to concentrate on transient flows caused by a sudden loading of sugar. First of all this gives us the possibility of observing dynamical behavior which allows us to compare quantitatively with theory and second, we can observe such dynamics using simple boundary conditions (e. g. closed ends), which are easily realized in experiments.

\subsection{Previous experimental work}
To study the osmotically driven flows, Eschrich, Evert and Young \cite[]{Eschrich:1972} conducted simple model experiments. Their setup consisted of a semipermeable membrane tube submerged in a water reservoir, modeling a phloem sieve element and the surrounding water-filled tissue.  At one end of the tube a solution of sugar, water and dye was introduced to mimic the sudden loading of sugar into a phloem sieve element. The motion of this ``sugar front" was monitored for different configurations of the tube by observing the motion of the dyed front. They conducted experiments with the tube closed at both ends (closed) and open at one end and closed at the other (semi-closed).

In the case of the closed tube, they found that the sugar front velocity decayed exponentially as it approached the far end of the tube. Also, they found the initial velocity of the sugar front to be proportional to concentration of the sugar solution. Through simple conservation arguments, which we shall go through briefly below, they showed that for a flow driven according to the pressure-flow hypothesis, the velocity of the sugar front is given by
\beq
u_f=\frac{L}{t_0}\exp \left(-\frac{t}{t_0}\right)\quad \text{where} \quad t_0=\frac{r}{2L_p \Pi},
\label{eq:esch1}
\eeq
where $t$ is time, $L$ is the length and $r$ is the radius of the tube, $L_p$ is the permeability of the membrane and $\Pi$ is the osmotic pressure of the sugar solution, equal to $RTc$ (where $R$ is the gas constant, $T$ the absolute temperature and $c$ the concentration in moles pr. volume) for dilute solutions of ideal molecules \cite[]{Landau:1980}. If one applies this result to the flow inside a single sieve element ($L=1$mm), one gets a characteristic velocity of $\simeq 7 \text{mh}^{-1}$, almost an order of magnitude larger than that observed in plants (see table \ref{table:phloem}).
The fact that equation \ref{eq:esch1} predicts a larger velocity is not surprising. One should keep in mind that the sieve cells consist of sieve elements separated by sieve plates and the resistance in strongly concentrated in the latter. Thus our experiments would model only the transport in a single sieve element without significantly changing the resistance. Secondly, the measurements in the table are supposed to be representative of a steady state situation and not the movement of a sugar front.

In their experiments Eschrich \etal found good qualitative agreement between their results and the prediction made by equation \ref{eq:esch1}.
 \begin{figure}
\centerline{\includegraphics[width=0.75\textwidth]{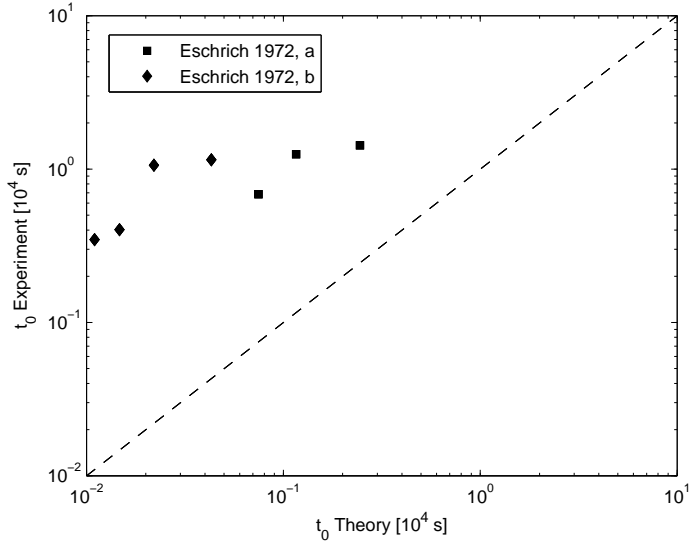}}
\caption{Experimental data from \cite[]{Eschrich:1972}. Data points marked with an a represents results from closed tube experiments and points marked with a b represents results from semi-closed experiments. $t_0^{exp}$ was taken from the original paper, figures 8 and 9. To calculate $t_0^{theory}$, the expression $t_0^{theory}=\frac{r}{2\L_p \Pi}$ was used with $r=3.5$mm and $L_p=3.2\times 10^{-12}$m(Pas)$^{-1}$. The values for $\Pi$ was found from figures 8 and 9.}
\label{fig:muncht0eschrich}
\end{figure}
However, as can be seen in figure \ref{fig:muncht0eschrich}, the quantitative agreement between experiments and theory was extremely poor, the theory generally predicting decay times at least an order of magnitude smaller than observed.
This discrepancy, however, is not surprising given the fact that Eschrich \etal did neither determine the fundamental properties of the membrane (permeability, elastic modulus), nor of the sugar (osmotic strength) independently. Also they did not take into account the unstirred concentration layers which may occur near the membrane walls, effectively lowering $\Pi$ \cite[]{Pedley:1980,Pedley:1981,Pedley:1983,Aldis:1988}. The disagreement between theory and experiment, and the fact that the tracking of the sugar front was done indirectly, leave the details of the observed process unclear. In particular, the experiments of Eschrich \etal did not include a continuous monitoring of the pressure. They state (p. 295) that the turgor pressure rapidly builds up to a constant value, as predicted theoretically. They observe, however, that the tubes start leaking after around 100-150 min. (typical running time of the experiments), and this seems to indicate that the pressure continues to grow. Another source of error could come from the fact that Eschrich \etal used a dye for tracking the sugar front instead of directly monitoring the sugar concentration. Finally, it is not clear whether the membranes are sufficiently impermeable to the sugar (sucrose) used.

To make progress on these issues we have refined the experiments done by Eschrich \etal to test the pressure-flow hypothesis more accurately. We have measured the permeability of the membrane tube, and the osmotic strength of the sugar solutions independently and we continuously monitor the pressure during the experiment.  Also, we have measured the elastic modulus of the membrane tube, to asses the importance of elastic effects in our system. For more detailed comparison with theory it is important to be able to assess the entire concentration profile, and to do this, we have introduced a new type of experiment, where the sugar concentration is determined directly by optical refraction.

\section{Experimental setup}
In our experiments, we used two setups. The first setup, from now on called setup I, is based on the design of Eschrich  \etal, but includes continuous pressure measurements. Further, to avoid leaking of sugar across the membrane, we use a sugar ({\it dextran}) with a much larger molecular weight than sucrose. The second setup, Setup II, was built in the spae of a prism to be able to detect the evolution of the entire sugar profile (again using {\it dextran}) rather than just the front position offered by previous experiments. The two setups are discussed in detail below. Details of the sugar and semipermeable membranes can be found in appendix \ref{app:setups}.
\subsection{Setup I}
\begin{figure}
\centerline{
\includegraphics[height=0.5\textheight]{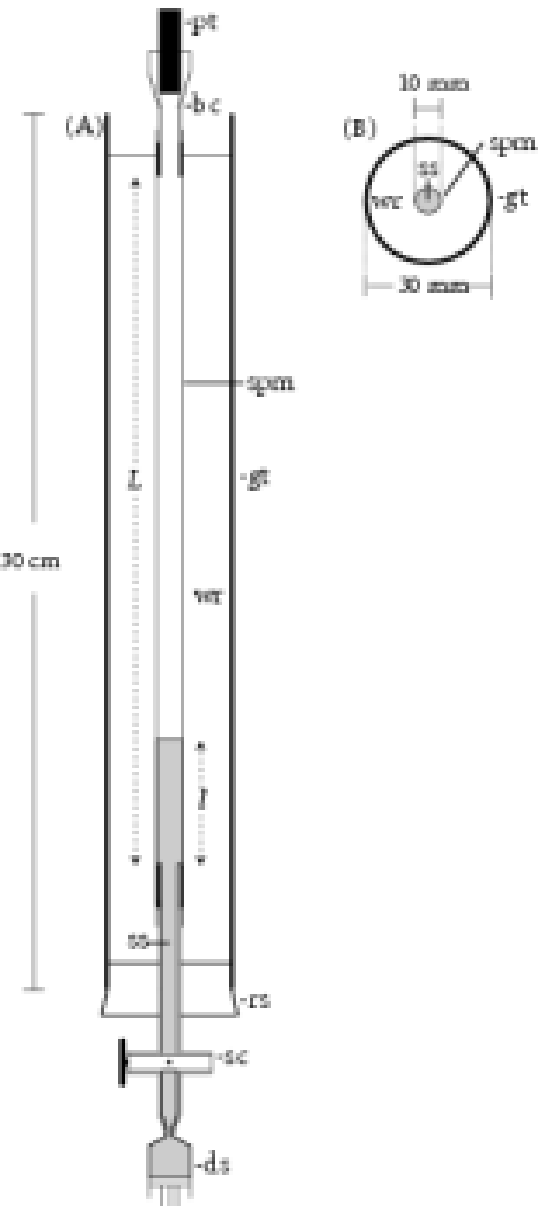}
\includegraphics[height=0.5\textheight]{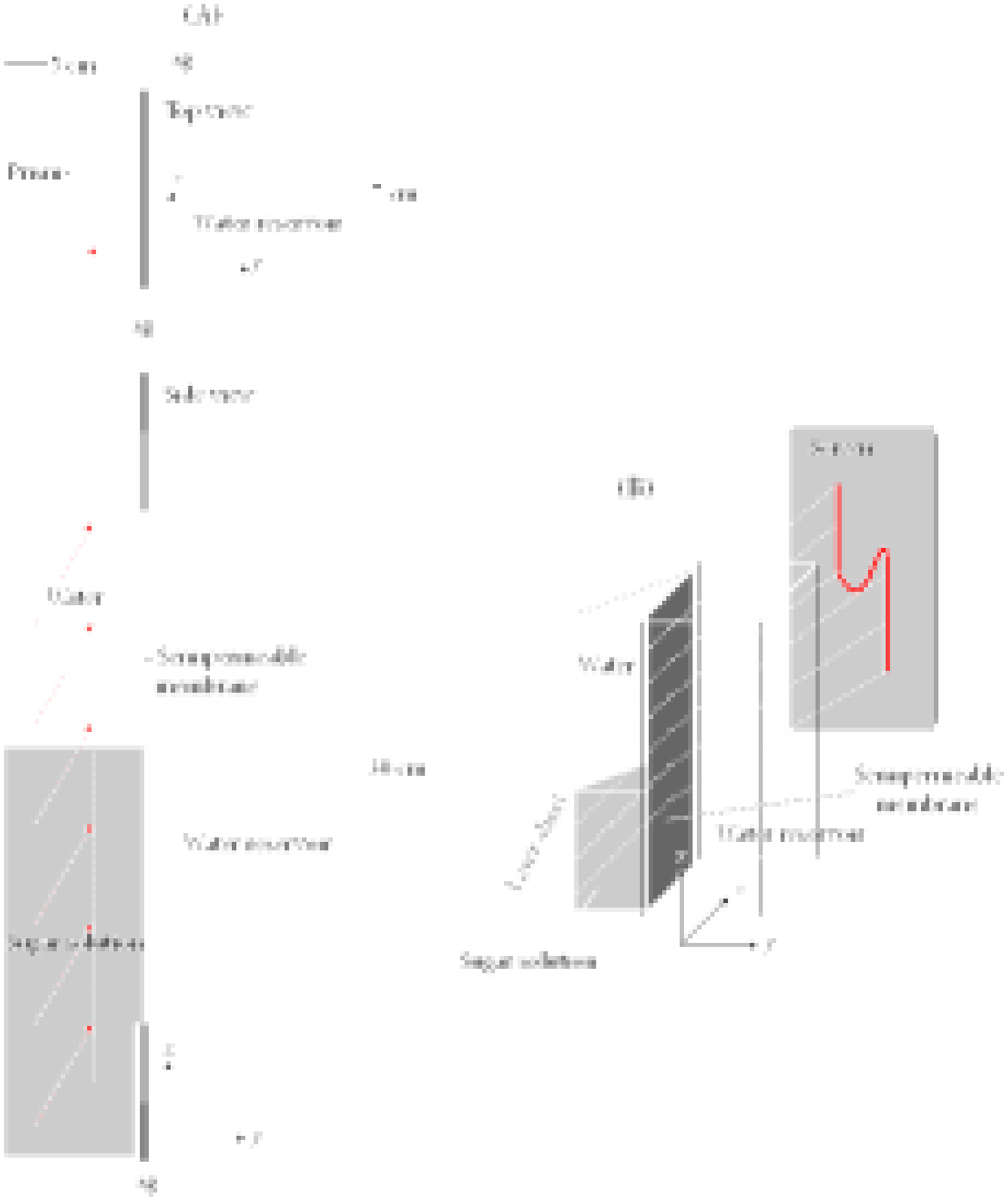}
}
\caption{Left: Setup I used to observe the movement of a sugar-dye solution (ss) inside a semipermeable membrane tube (spm). $L$ length of membrane tube; $l$ initial sugar front height; ds disposable syringe; gt glass tube; rs rubber stopper; sc stopcock; wr water reservoir; bc brass cylinder; pt pressure transducer.
\newline Right: Setup II. See details in the text.}
\label{fig:setups}
\end{figure}
Setup I, shown on the left in figure \ref{fig:setups}, consisted of a $30$ cm long, $30$ mm wide glass tube in which a semipermeable membrane tube of equal length and a diameter of $10$ mm was inserted. At one end, the membrane tube was fitted over a glass stopcock equipped with a rubber stoppper. At the other end, the membrane tube was fitted over a brass cylinder equipped with holder to accommodate a pressure transducer for measuring the pressure inside the membrane tube.

After filling the $30$ mm wide glass tube with water, water was pressed into the semipermeable tube with the syringe. Care was taken that no air bubbles were stuck inside the tube. For introducing the sugar solution into the tube, a syringe was filled with the solution and then attached to the lower end of the stopcock which was kept closed. After fitting the syringe, the stopcock was opened and the syringe piston was very slowly pressed in, until a suitable  part of the tube had been filled with the solution. Care was taken to avoid any mixing between the sugar solution and the water already present in the semipermeable tube.
To track the movement of the sugar solution it was mixed with a red dye and data was recorded by taking pictures of the membrane tube at intervals of 15 minutes using a digital camera. Details of how the motion of the sugar front was derived from the images is discussed in appendix \ref{app:setupI}.

\subsection{Setup II}
Setup II, shown on the right in figure \ref{fig:setups} consisted of a hollow isosceles glass prism and a Plexiglas cuboid in osmotic contact through a membrane. The glass prism was fitted with a pressure transducer for measuring the pressure inside the membrane tube.

When preparing an experiment, a piece of membrane was fitted in a narrow gap between the prism and the cuboid. The prism was then filled to a suitable height with a sugar solution and pure water was carefully deposited on top of the sugar solution to create a sharp sugar front.
Then, the cuboid was filled with water, and the pressure transducer was mounted, thereby closing the prism.

To track the time evolution of the the sugar front inside the prism, we used the refraction of a laser sheet passing through it.  The laser sheet was generated by shining a laser beam, generated by a Melles Griot 3.1 mW laser, through a glass rod.
When passing through the prism, light would deviate depending on the local index of refraction. The index of refraction varies linearly with sugar concentration and thus by looking at the refracted laser sheet projected onto a screen, we were able to reconstruct the concentration profile inside the prism.
A camera recorded images of the screen at regular intervals to track the moving concentration profile. The procedure used for obtaining the concentration profile inside the prism from the images aquired is discussed in appendix \ref{app:setupII}
\section{Experimental results}
\label{sec:results}
\subsection{Experimental results, Setup I}
\label{sec:setupIresults}
\begin{figure}
\centerline{\includegraphics[width=1.2\textwidth]{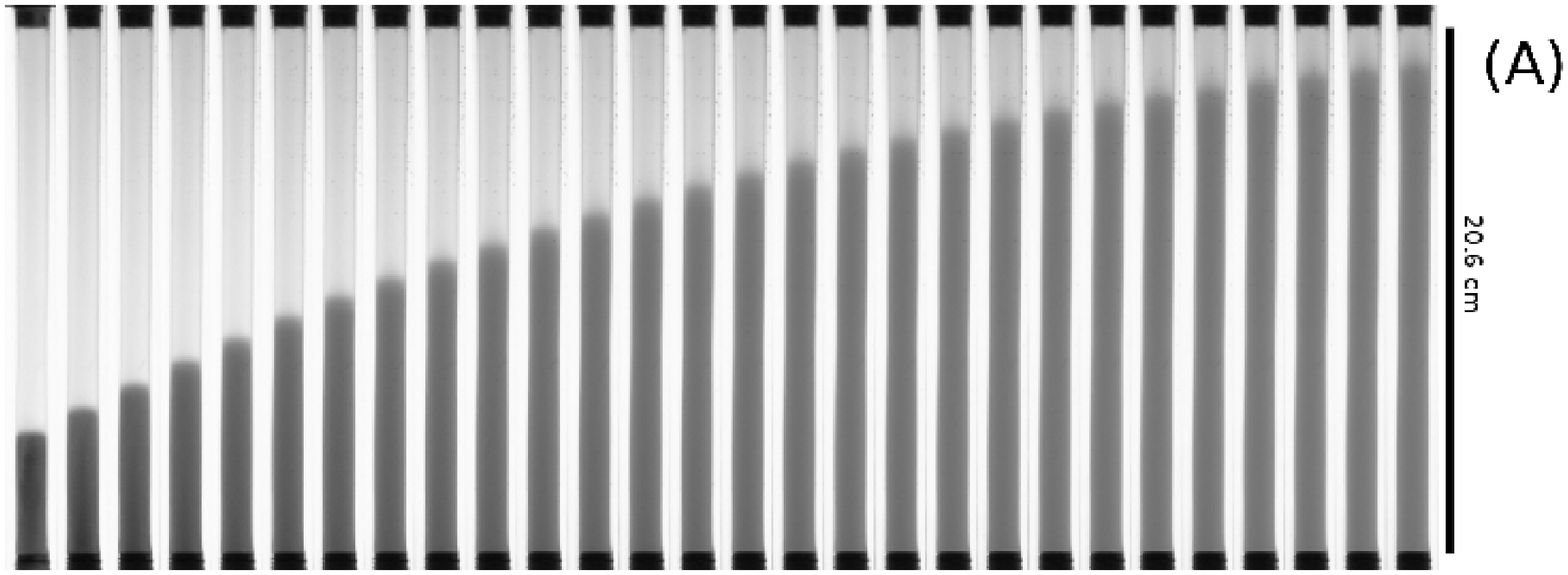}}
\centerline{\includegraphics[width=0.55\textwidth]{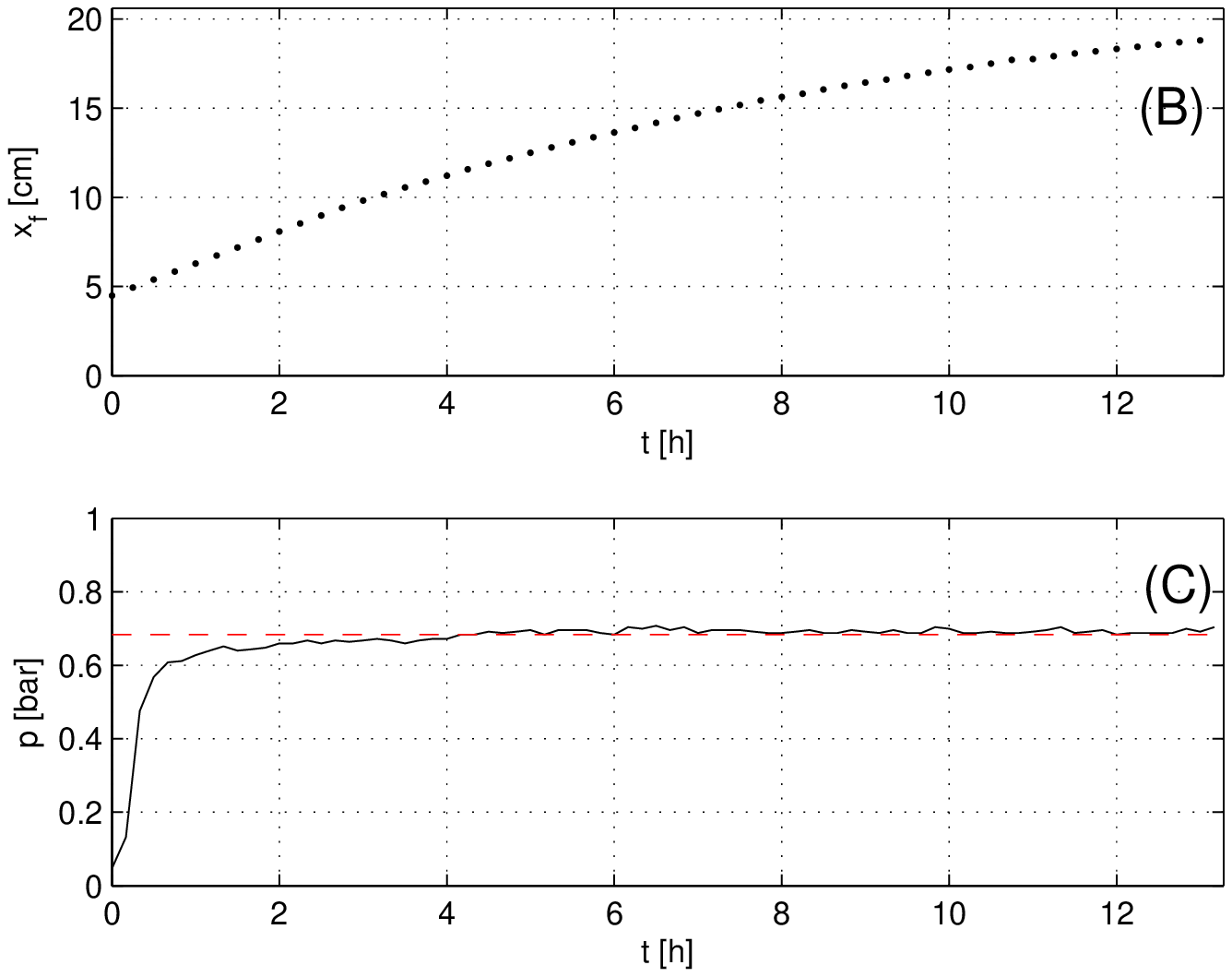}\includegraphics[width=0.55\textwidth]{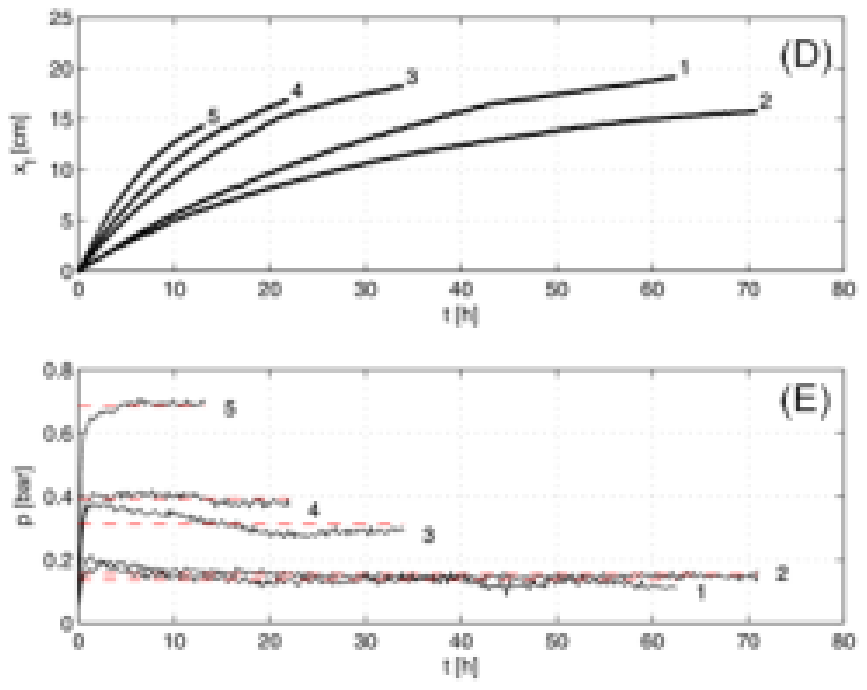}}
\caption{Experimental results from setup I. Top: Time series of pictures taken in experiment 5. Time increases from left to right in steps of 30 minutes. See details of the sugar solutions used in table \ref{table:setupIresults}. Middle, left: Plot of the front position versus time obtained from the images above. Bottom, left: Plot of the pressure inside the tube versus time. The (red) dashed line is the osmotic pressure of the solution, taken to be the average value of the pressure from $t=2$ h until the end of the experiment. Middle, right: Plots of the sugar front position versus time for different sugar concentrations, as indicated in table \ref{table:setupIresults} Bottom, right: Plots of the pressure inside the membrane tube for different sugar concentrations.}
\label{fig:setupIresults}
\end{figure}
\begin{table}
\begin{center}
\begin{tabular}{ l  c  c  c  c  c }
              & 1                          & 2          & 3                  & 4                  & 5 \\[3pt]
Mean sugar concentration, $\bar c$ [mM]   & 1.5$\pm$0.3           &2.10$\pm$0.03  &2.4$\pm$0.2    &4.2$\pm$0.7    &6.8$\pm$0.1 \\
Osmotic pressure, $\Pi$ [bar]           &0.14 $\pm$ 0.02        &0.15$\pm$0.01  &0.31$\pm$0.03  &0.39$\pm$0.01  &0.68$\pm$0.02\\
Membrane tube length, $L$ [cm]              &28.5                   & 20.8          &28.5           &28.5           & 20.6          \\
Initial front height, $l$ [cm]              &4.9                    & 3.7           &6.6            &6.5            & 4.8           \\
\end{tabular}
\end{center}
\caption{Data for the experimental runs shown in figure \ref{fig:setupIresults}.}
\label{table:setupIresults}
\end{table}
The motion of the sugar front was investigated for solutions of varying sugar concentration.
An example of a set of data is shown in figure \ref{fig:setupIresults}. In (A) are the raw images, which after processing gives (B) showing the position of the sugar front, $x_f$, as a function of time. The errorbars on $x_f$ are estimated to be $\pm 1$ mm, but are too small to be seen.
Finally, (C) shows the pressure inside the tube as a function of time.
At first, a linear motion of the front is observed with a front velocity of $\sim 1$ cm/h. This is then followed by a decrease in the front velocity as the front approaches the end of the tube.
The pressure is seen to rise rapidly during the first hour before settling to a constant value, indicated by a red, dashed line. This constant value is taken to be the osmotic pressure, $\Pi$, of the sugar solution.
Looking at (A), one observes that diffusion has the effect of dispersing the front slightly as time passes.
Below the front, the concentration seems to be uniform throughout the cross-section of the tube, and there is no indication of large boundary layers forming near the membrane walls.

Similar experiments with different sugar concentrations were made and a plot of the results can be seen on the right in figure \ref{fig:setupIresults} (D) and (E). Qualitatively both the motion of the front and of the pressure follow the same pattern as shown on the left. One notices that the speed with which the fronts move are related to the mean sugar concentration inside the membrane tube, with the high concentration solutions moving faster than the low concentration ones. The reason why $2$ seems to be moving slower than $1$ is that experiment $2$ was conducted in a slightly shorter membrane tube than $1$, thereby decreasing the characteristic velocity as we shall see later.
\subsection{Experimental results, Setup II}
\begin{figure}
\centerline{
\includegraphics[width=0.8\textwidth]{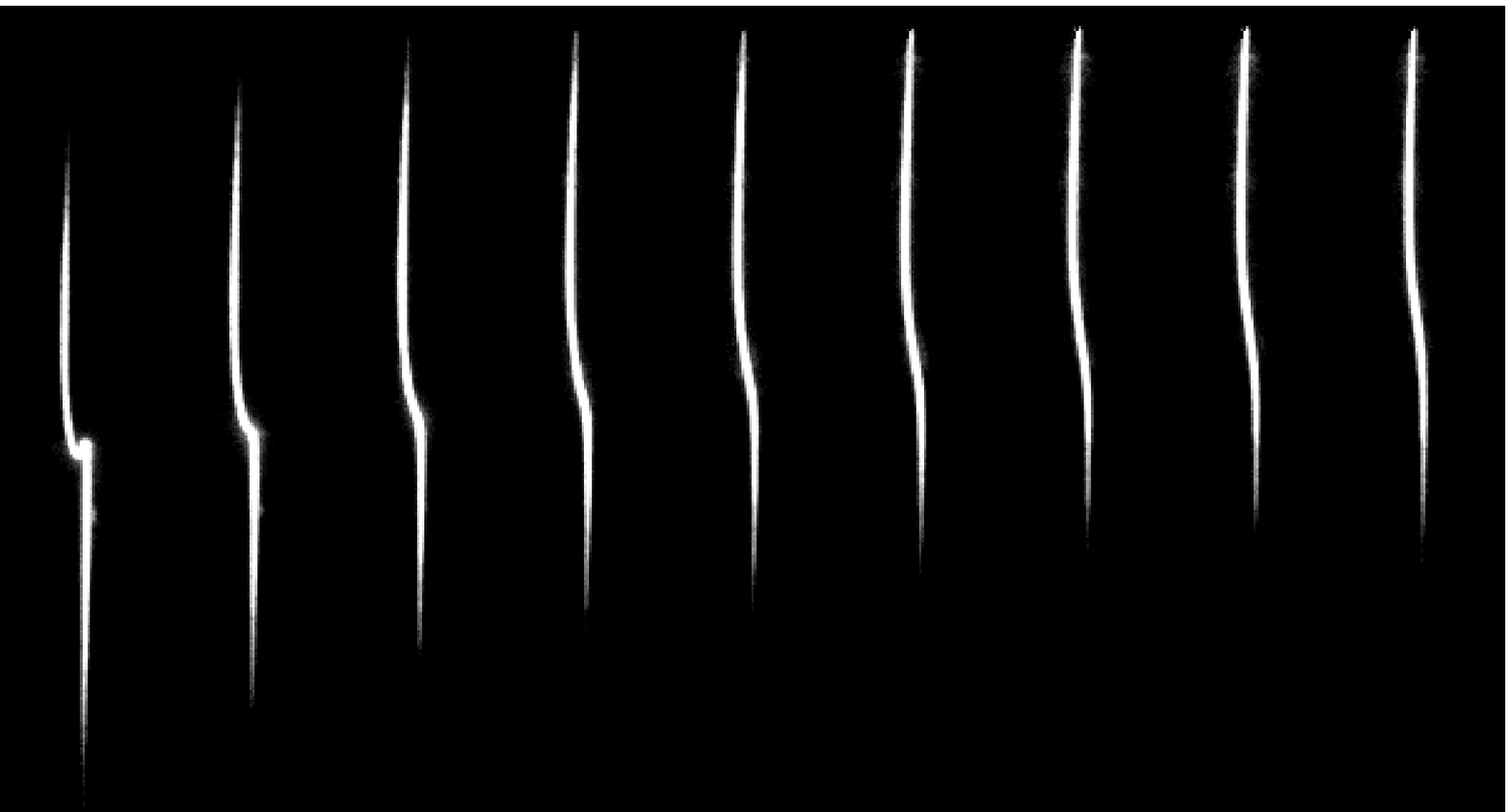}}
\centerline{\includegraphics[width=1\textwidth]{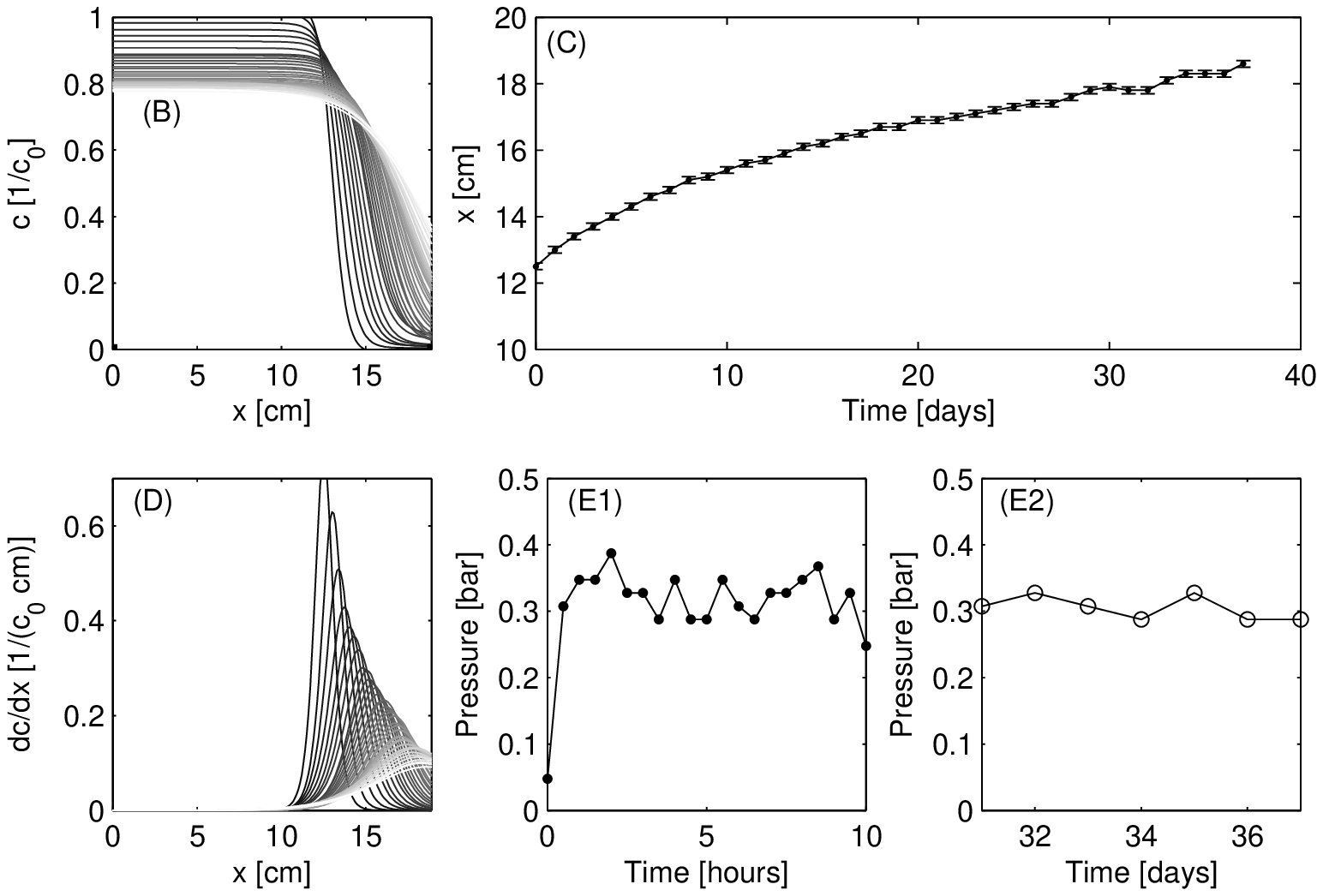}}
\caption{Results from setup II. In (A) the raw data images are shown. In (B) the concentration profile extracted from (A) is shown. (C) shows the front position extracted from (B) by finding the maximum of the concentration gradient, shown in (D). Finally, (E, 1-2) shows the pressure inside the prism. The dashed line indicates that the pressure sensor was accidentaly off-line. The data shows that the pressure rather quickly (within a few hours) reaches a constant level corresponding to the osmotic pressure of the sugar solution (See appendix A 2.2.)}
\label{fig:riocomp1}
\end{figure}
Figure \ref{fig:riocomp1} shows the data collected using setup II. At the top, a time series of pictures is depicted showing the refracted laser-light projected onto a screen, the time between each image being one day. Comparing the upper and lower parts of each picture,  one generally observes a deflection to the right at the bottom, corresponding to a high sugar concentration at the bottom of the prism. In the intermediate region one sees a dip in the refracted light, corresponding to a strong concentration gradient. The dip gradually flattens while it advances upwards, representing a sugar front which advances while it broadens.
This process can be seen directly on the left in the middle row, which shows the time evolution of the sugar concentration obtained from the images. Starting from a steep concentration profile, we see that the front moves forward while it flattens. On the left in the bottom row, the time evolution of the concentration gradient is depicted, clearly showing a peak which broadens while it moves forward. Finally, on the right, the position of the sugar front and the pressure inside the prism is plotted as a function of time is shown. The errorbars on $x_f$ are $\pm 1$ mm, found as discussed below.
\subsubsection{The effects of diffusion}
\begin{figure}
\centerline{
\includegraphics[width=1\textwidth]{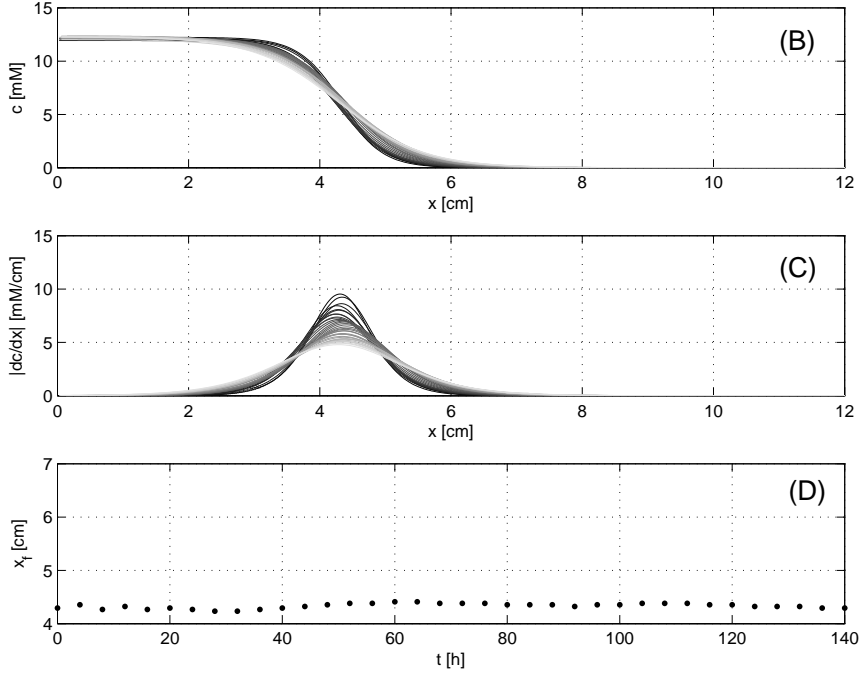}
}
\caption{Experimental results from diffusion experiments made with setup II.}
\label{fig:expres2}
\end{figure}

To study the effects of diffusion on the dynamics of the sugar front separately, an experiment was made with setup II, in which the membrane separating the two compartments were removed. The experiment was then prepared in the usual way, and the motion of the front recorded. The results of this is shown on the right in figure \ref{fig:expres2}. Starting from a steep concentration gradient, we observe that the front flattens but otherwise does not move much.

Comparing figures \ref{fig:setupIresults} and \ref{fig:riocomp1} we observe, that while the front moves $\tilde 2$ cm due to osmosis in $72$ hours, it does not move at all in $140$ hours due to diffusion. Thus, while diffusion has a flattening effect, it plays little role in the forward motion of the front.

Since the front did not move due to diffusion, the fluctuations in the front position seen in figure \ref{fig:expres2} (D) gives a measure of the uncertainity of a single measurement of the front position. Taking the standard deviation of the fluctuations gives an uncertainty of $\pm 1$ mm, shown as errorbars in figure \ref{fig:riocomp1} (C).
\section{Theoretical analysis}
\subsection{Front propagation}
Before moving on to a more thorough mathematical analysis we shall, following the analysis made by \cite[]{Eschrich:1972}, show that the motion of the sugar front can be understood through simple conservation arguments. To that end, let us consider the situation in setup I. Let $x_f$ denote the position of the sugar-dye front, and let $V_1$ denote the volume behind the front and $V_2$ that ahead of it. Taking the tube to be inelastic and the fluid inside incompressible, we must have that $\frac{dV_1}{dt}=-\frac{dV_2}{dt}$. If we let $p(x)$ and $p_0$ denote the hydrostatic pressure inside and outside the membrane tube and $c(x)$ the concentration averaged over the cross-section of the tube at position $x$, the volume flux across a unit area of the membrane is
\beq
J(x)=L_p\left(p_0-p(x)+RTc(x)\right).
\eeq
Any hydrostatic pressure gradient inside the tube will occur only due to viscous flow, but for large tubes and slow flows, this effect is entirely negligible. Thus, since
\beq
\int_0^L J(x')dx'=0
\eeq
we get, that
\beq
p-p_0=\frac{RT}{L}\int_0^L c(x')dx'\equiv RT\bar c
\eeq
The rate of change of volume 2 is then
\beq
\frac{dV_2}{dt}=2\pi r L_p\int_{x_f}^L J(x')dx' =-2\pi rL_p RT\bar c(L-x_f)
\eeq
Finally, using that
\beq
\frac{dV_2}{dt}=-\pi r^2 \frac{dx_f}{dt}
\eeq
we get for the front position
\beq
\frac{dx_f}{dt}=\frac{2L_pRT\bar c}{r}(L-x_f)=\frac{1}{\tau}(L-x_f)
\eeq
which has the solution
\beq
    x_f(t)=L-(L-l)\exp\left( -\frac t{\tau}\right)
\eeq
where $l=   x_f(0)$ is the position of the front at $t=0$.
This simple result, shows that the relaxation time
\beq
	\tau=\frac{r}{2L_pRT\bar{c}}
	\label{eq:tau}
\eeq
for the front propagation depends only on three quantities; the membrane permeability, the osmotic pressure of the sugar solution and the ratio of membrane circumference to cross-section area.

The neglect of elastic deformations of the tube is appropriate for the experiments made in both setups, as shown in Appendix A.2, as long as the pressures remain below around 1.2 bars.  Since we use the large dextran molecules, the osmotic pressures remain well below this limit. When sucrose is used, pressures become much larger and elastic properties can become important. In Appendix E, we show how the above results would change, if elastic properties are taken into account.

\subsection{Derivation of the flow equations}
To formalize the assumptions made above, we will now derive the equations of motion for osmotically driven flows, with the geometry of setup I in mind. In appendix \ref{sec:othergeom}, we shall see that under certain conditions the equations are also valid in other geometries, such as the triangular geometry of setup II.

The equations of motion for osmotically driven flows have been derived and analyzed thoroughly several times in the literature \cite[see eg. ][]{Weir:1981} and have been studied carefully numerically, \cite[]{Thompson:2003,Thompson:2003b,Henton:2002}.
For the sake of completeness, we shall include a short derivation of these.

\begin{figure}
\centerline{\includegraphics[width=1\textwidth]{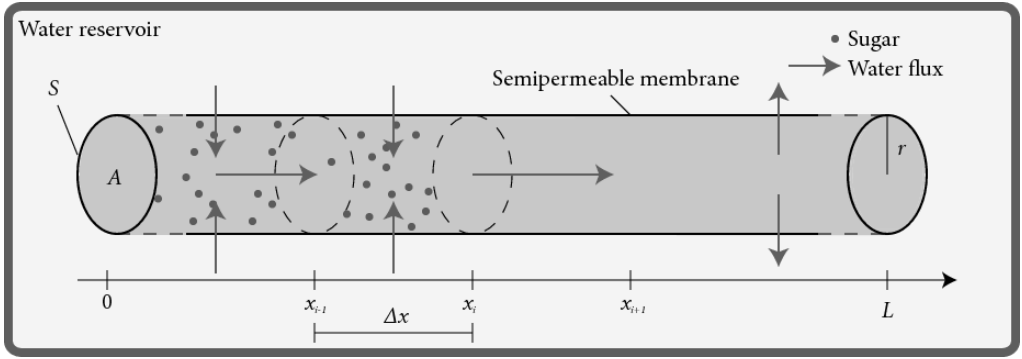}}
\caption{Sketch of the tube.}
\label{fig:tube1}
\end{figure}

We consider a tube of length $L$ and radius $r$, as shown in figure (\ref{fig:tube1}). The tube has a constant cross section of area $A=\pi r^2$ and circumference $S=2\pi r$ and its walls are made of a semipermeable membrane with permeability $L_p$. Inside the tube is a solution of sugar and water with concentration $c(x)$. Throughout this paper, we shall study the transient dynamics generated by an asymmetrical initial concentration distribution, where the sugar is initially localised to one end of the tube with a concentration level $c_0$. The tube is surrounded by a water reservoir, modelling the water surrounding the membrane tube in setup I.

We shall assume that $L\gg r$ and that the radial component of the flow velocity inside the tube is much smaller than the axial component, as is indeed the case in the experiments. With these assumptions, we will model the flow in the spirit of lubrication theory and consider only a single average axial velocity component $u(x,t)$. Also, we will assume that the concentration, $c$, is independent of the radial position, $\rho$, an assumption that can be verified experimentally in setup II.
%

%
Let us now consider the equation of volume conservation by looking at a small section of tube between $x_{i-1}$ and $x_i$. The volume flux into the section due to advection is
\beq
    A(u_{i-1}-u_i),
\label{eq:floweq4}
\eeq
where the axial flow velocities are taken to be $u_{i-1}$ and $u_i$ at $x_{i-1}$ and $x_i$, respectively.
The volume flux inwards across the membrane due to osmosis \cite[see eg.][]{Schultz:1980} is
\beq
    S\Delta xL_p(RTc-p).
\label{eq:floweq5}
\eeq
where $p$ is the pressure. For clarity we use the van't Hoff value $\Pi = RTc$ for the osmotic pressure, which is only valid for ideal solutions. When comparing with experiments we replace this by $\Pi = \gamma c$, where $\gamma$ must be experimentally determined (see appendix A.2), and thus only use that $\Pi$ is linear in $c$ for small concentrations.
Assuming conservation of volume, we get that
\beq
    A(u_{i-1}-u_i)+S\Delta xL_p(RTc-p)=0
\label{eq:floweq6}
\eeq
Letting $\Delta x\to 0$ and using that the cross-section to perimeter ratio reduces to $\frac r2$, this becomes
\beq
    \frac r2\partiel ux=L_p(RTc-p)
\label{eq:floweq8}
\eeq
For these very slow and slowly varying flows, the time dependence of the Navier-Stokes equation can be neglected and the velocity field is determined by the instantaneous pressure gradient though the Poiseuille or d'Arcy relation (for a circular tube)
\beq
    u=-\frac{r^2}{8\eta}\partiel px,
\label{eq:floweq9}
\eeq
where $\eta$ is the dynamic viscosity of the solution, typically $\sim 1.5\times 10^{-3}$ Pa s in our experiments.

Differentiating equation (\ref{eq:floweq8}) with respect to $x$ and inserting the result from equation (\ref{eq:floweq9}) we get for the conservation of water that
\beq
    RT\partiel cx=\frac {r}{2L_p} \ppartiel ux-\frac{8\eta}{r^2}u.
\label{eq:jfm1}
\eeq
The final equation expresses the conservation of sugar advected with veolcity $u$ and diffusing with molecular diffusivity $D$
\beq
    \partiel ct +\partiel{uc}x=D\ppartiel cx.
\label{eq:jfm2}
\eeq
The set of equations \ref{eq:jfm1}--\ref{eq:jfm2} are equivalent to those of Thompson and  Holbrook \cite[]{Thompson:2003b} except for the fact that we have removed the pressure by substitution, and that we do not consider elastic deformations of the tube.
\subsubsection{Non-dimensionalization of the flow equations}
To non-dimensionalize equations (\ref{eq:jfm1}) and (\ref{eq:jfm2}), we introduce the following scaling
$$
    c=c_0C,\qquad u=u_0U, \qquad x=LX, \qquad t=t_0\tau.
    \label{eq:scaling}
$$
$L$ has been chosen such that the spatial domain is now the unit interval $X\in [0,1]$, $u_0=L/t_0$ and $c_0$ is the initial concentration level in one end of the tube. Choosing further
\beq
t_0=\frac{r}{2L_pRTc_0}, \quad M=\frac{16\eta  L^2L_p}{r^3} \quad\textrm{and}\quad \bar D=\frac{D}{u_0 L}=\frac{Dr}{2RTc_0L^2L_p},
\eeq
\label{page:u0}
and inserting in equations \ref{eq:jfm1} and \ref{eq:jfm2}, we get the non-dimensional flow equations.
\beq
    \ppartiel UX -MU=\partiel CX,
\label{eq:volumeconservationnondim}
\eeq
\beq
    \partiel C\tau+\partiel{UC}{X}=\bar{D}\ppartiel CX.
\label{eq:soluteconservationnondim}
\eeq
Going back to the original notation
$$
    X\to x,\qquad U\to u,\qquad C\to c,\qquad  \tau\to t
$$
we finally obtain
\beq
    \ppartiel ux -Mu=\partiel cx,
\label{eq:volumeconservationnondim2}
\eeq
\beq
    \partiel ct+\partiel{uc}{x}=\bar{D}\ppartiel cx.
\label{eq:soluteconservationnondim2}
\eeq
\begin{table}
\begin{center}
\begin{tabular}{lcc}
 					&$M$			&$\bar D$\\[3pt]
Setup I  				&  $2\times 10^{-8}$	&$6\times 10^{-5}$\\
Setup II 				&$10^{-9}$ 		&$2\times 10^{-2}$ \\
Single Sieve element ($L=1$ mm)  	&$5\times 10^{-4}$	 &$5\times 10^{-4}$ \\
Leaf ($L=1$ cm) 			&$5\times 10^{-2}$ 	&$5\times 10^{-5}$ \\
Branch ($L=1$ m) 			&$5\times 10^{2}$	 &$5\times 10^{-7}$ \\
Small tree ($L=10$ m)			& $5\times 10^{4}$ 	&$5\times 10^{-8}$ \\
\end{tabular}
\end{center}
\caption{Values of the parameters $M$ and $\bar D$ in various situations.}
\label{table:mandd}
\norule
\end{table}
 \followon
\begin{table}
\begin{center}
\begin{tabular}{llc}
Quantity 					& Magnitude 						& Reference.\\[3pt]
Radius [$\mu$m]					& 4.5 (Fava bean), 4 (Winter squash), 6--25 		&$\dagger$, $*$, $||$\\
Length [mm]					& 0.09 (Fava bean), 0.1--3 				&$\dagger$, $||$\\
  Flow velocity [mh$^{-1}$]			&0.5--1, 0.2--2 					&$\dagger$, $||$\\
  Elastic Modulus [MPa] 			&17, 5.6--7.4 (Ash)					& $\bullet$, $\langle$\\
  Permeability [$10^{-11}$ ms$^{-1}$Pa$^{-1}$]	&5,1.1 (Zitella translucence) 				&$\bullet$, $\rangle$\\
  Sucrose concentration [M] 			&0.3--0.9 						& $*$\\
\end{tabular}
\end{center}
\caption{Characteristic properties of phloem sieve elements. References: $\dagger$ \cite{Knoblauch:1998}, $*$ \cite{Taiz:2002}, $||$ \cite{Nobel:1999}, $\bullet$ \cite{Thompson:2003}, $\langle$ \cite{Niklas:1992}, $\rangle$ \cite{Eschrich:1972}.}
\label{table:phloem}
\end{table}
The parameter $M$ corresponds to the ratio of axial to membrane flow resistance, which we shall refer to as the {\it M\"unch  number}. This is identical to the parameter $\hat{F}$ in \cite{Thompson:2003b}. The second parameter $\bar D$ is the ratio of diffusive and advective solute flux. Thus, the longer the tube is the less important diffusion becomes and the more important the pressure gradient due to viscous effects become.


Values of the parameters $M$ and $\bar D$ in different situations can be seen in table \ref{table:mandd}.
The typical magnitude of the parameters $M$ and $\bar D$ in plants are found from the values also given in table \ref{table:mandd}:
$$
    r=10\;\mu\text m,\qquad \eta=1.5\times 10^{-3}\;\text{Pas},\qquad u_0=2\;\text{mh}^{-1},\qquad L_p=2\times 10^{-11}\text{m(Pas)}^{-1}.
$$
We observe, that $M$ and $\bar D$ are small in both experiments, and that for short distance transport in plants this is also the case. However, over length-scales comparable to a branch ($L=1$ m) or a small tree ($L=10$ m) $M$ is large, so in this case the pressure gradient is not negligible.

When deriving the equations for osmotically driven flows, we have
assumed that the concentration inside the tube was a function of $x$ and $t$
only. However, the real concentration inside the tube will also depend
on the radial position $\rho$ in the form of a concentration boundary layer near
the membrane, in the literature called an {\em unstirred layer} \cite[]{Pedley:1983}
Close to the membrane, the concentration $c_m $ is lowered compared to the bulk
value, $c_b$, because sugar is advected away from the membrane by the influx of water.
%
%
This, in turn, results in a lower influx of water, ultimately causing the axial flow inside the tube to be slower than expected. In our experiments we see no signs of such boundary layers and apparently their width and the effects on the bulk flow are very small.

\section{Solution of the flow-equations}
The equations governing the time evolution of a sharp sugar front has been known for the closed and semi-closed tube since the work of Eschrich \etal . However, their solutions yield only the position of the front, and not the concentration profiles in front of and behind the concentration front. Analytic solutions giving the time evolution of the entire concentration profile has been found for the closed tube for $M=\bar D=0$ by G. J. Weir \cite[]{Weir:1981} and by H. L. Frisch \cite[]{Frisch:1976} for the the semi-closed tube for $M=0,$ $\bar D\neq 0$. In both cases the authors have started from piecewise constant Heaviside-like initial concentration profile. To extend this work, we shall present a method for solving the equations of motions analytically using Riemann's method of characteristics. For an arbitrary initial condition, this method will generally yield an implicit solution. Only in special cases will it yield a closed formula for the solution. The method works for $M=\bar D=0$ and for closed and semi-closed tube geometries.

For arbitrary values of $M$ and $\bar D$, we cannot solve the equations of motion analytically and thus have to use numerical methods. This topic has been the focus of much work both in the steady-state case \cite[]{Thompson:2003} and in the transient case \cite[]{Henton:2002}. However, no formulation capable of handling all different boundary conditions has so far been presented. Therefore, we show that using Green's functions, the equations of motion can be transformed into a single integro-differential equation, which can be solved using standard numerical methods.
\subsection{Results for small M\"unch number}
\label{sec:analsol}
In the limit $M=\bar{D}=0$ the equations become
\beq
    \ppartiel ux=\partiel cx,
\label{eq:rie1}
\eeq
\beq
    \partiel ct+\partiel{uc}{x}=0.
\label{eq:rie2}
\eeq
By integrating equation equation \ref{eq:rie1} with respect to $x$, we get that
\beq
    \partiel ux = c+F(t).
\label{eq:rie3}
\eeq
If we choose $u(0)=u(1)=0$, $F(t)$ becomes
\beq
    F(t)=-\int_0^1 c\ud x\equiv-\bar{c}(t),
\label{eq:rie4}
\eeq
Using \ref{eq:rie3} in equation \ref{eq:rie2} gives
\beq
    \partiel{}{x}\left(\partiel ut +u\left(\partiel ux +\bar{c}\right)\right)=0.
\eeq
Integrating with respect to $x$ and using the boundary conditions on $u$, this becomes
\beq
    \partiel ut +u\partiel ux=-\bar{c}u.
\label{eq:burgers1}
\eeq
Equation \ref{eq:burgers1} is a damped Burgers equation \cite[see eg.][]{Gurbatov:1991}, which can be solved using Riemann's method of characteristics. The characteristic equations are
\begin{eqnarray}
    \hard ut&=&-\bar{c}u \quad \label{eq:chareq1}\\
    \hard xt &=& u\label{eq:chareq2}.
\end{eqnarray}
Equation \ref{eq:chareq1} has the solution
\beq
    u=u_0(\xi)\exp(-\bar{c}t),
\label{eq:chareq3}
\eeq
where the parametrization $\xi(x,t)$ of the initial velocity has to be found from
\beq
    x=\xi+\frac{1}{\bar{c}}u_0(\xi)\left(1-\exp(-\bar{c}t)\right)
\label{eq:chareq4}
\eeq
where $\xi=x$ at $t=0$.
\subsubsection{Solution for piecewise constant initial concentration}
\begin{figure}
\centering
\centerline{\includegraphics[width=1\textwidth]{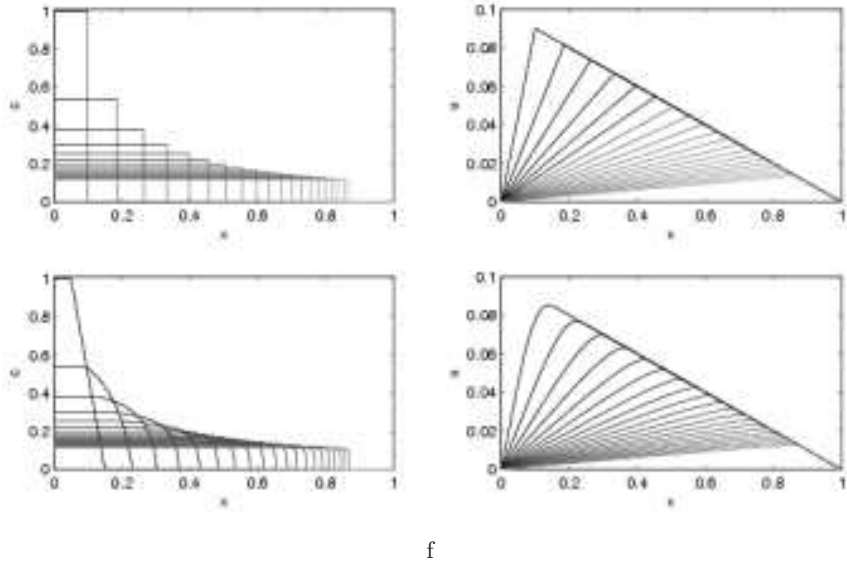}}f
\caption{Top: Plot of the analytical solution for a piecewise constant initial concentration. $\lambda=0.1$, $c_I=1$ and $\bar c=0.1$. Bottom: Plot of the analytical solution for a piecewise linear initial concentration. $\lambda_1=0.05$, $\lambda_2=0.15$, $c_I=1$ and $\bar c=0.1$. Time increases from black to gray in steps of one unit of time.}
\label{fig:analsolu1}
\end{figure}

To be able compare our method to the results obtained by \cite[]{Weir:1981}, we will use a Heaviside step function as initial condition on $c$
\beq
c(x,t=0)=c_I H(\lambda-x)=
    \begin{cases}
    c_I & \text{for}\quad 0\leq x\leq\lambda,\\
    0 & \text{for}  \quad \lambda < x\leq 1,
    \end{cases}
\label{eq:rie5}
\eeq
Equation \ref{eq:rie3} now enables us to find the initial condition on the velocity
\begin{eqnarray}
    u(x,t=0)&=&\int_0^x(c(x',0)-\bar{c})\ud x'=\int_0^x(c(x',0)-\lambda c_I)\ud x'\\
    &=&
    \begin{cases}
    (c_I-\bar{c})x & \text{for}\quad 0\leq x\leq\lambda,\\
    \bar{c}(1-x) & \text{for}  \quad \lambda < x\leq 1,
\label{eq:initialu}
\end{cases}
\end{eqnarray}
From equation \ref{eq:initialu}, we have that
\beq
    u_0(\xi)=
    \begin{cases}
    (c_I-\bar{c})\xi & \text{for}\quad 0\leq \xi\leq\lambda,\\
    \bar{c}(1-\xi) & \text{for}  \quad \lambda < \xi\leq 1.
\label{eq:chareq5}
\end{cases}
\eeq
Then, solving for $\xi(x,t)$ in equation \ref{eq:chareq4} gives
\beq
    \xi(x,t)=
\begin{cases}
    \frac{x}{1+\frac 1\lambda(1-\lambda)(1-\exp(-\bar{c}t)}&\quad\text{for}\quad x\in I_1\\
    \frac{x-1+\exp(-\bar{c}t)}{\exp(-\bar{c}t)}&\quad\text{for}\quad x\in I_2
\label{eq:chareq6}
\end{cases}
\eeq
where the intervals $I_1$ and $I_2$ are defined by
\beq
    I_1=[0,1-(1-\lambda)\exp(-\bar{c}t)],
\label{eq:chareq7}
\eeq
\beq
    I_2=[1-(1-\lambda)\exp(-\bar{c}t),1].
\label{eq:chareq8}
\eeq
Finally, $u(x,t)$ is calculated from equation \ref{eq:chareq3}
\beq
    u(x,t)=
    \begin{cases}
    \frac{(c_I-\bar{c})\exp(-\bar{c}t)x}{\frac 1\lambda(1-\lambda)(1-\exp(-\bar{c}t))}, & \quad\text{for}\quad x\in I_1\\
    \bar{c}(1-x),&\quad\text{for}\quad x\in I_2
\label{eq:chareq9}
\end{cases}
\eeq
which is equivalent to the result obtained by \cite[]{Weir:1981}. The solution is plotted in figure \ref{fig:analsolu1}, top.
We can now calculate the instantaneous sugar front position $x_f$ and velocity $u_f$ using the right boundary of $I_1$ from equation \ref{eq:chareq7}
\beq
    x_f(t)=1-(1-\lambda)\exp(-\bar{c}t),
\label{eq:chareq10}
\eeq
\beq
    u_f(t)=\hard{x_f}t=\bar{c}(1-\lambda)\exp(-\bar{c}t).
\label{eq:chareq11}
\eeq
Similarly, $c(x,t)$ is given by
\beq
    c(x,t)=\frac{\bar{c}}{1-(1-\lambda)\exp(-\bar{c}t)}H(x_f-x).
\label{eq:chareq12}
\eeq
Going back to dimensional variables, equations (\ref{eq:chareq10}) and (\ref{eq:chareq11}) become
\beq
    x_f(t)=L-(L-l)\exp\left( -\frac t\tau\right) \quad \text{and}
\label{eq:chareq13}
\eeq
\beq
    u_f(t)=\frac{L}{\tau}\exp\left( -\frac t{\tau}\right),
\eeq
where $L$ is the length of the membrane tube, $l$ is the initial front position and the decay-time $\tau$ is accordance with the simple argument leading to equation (\ref{eq:tau}).

%
%
\subsubsection{Solution for piecewise linear initial concentration}
As initial condition on $c$, we will use the piecewise linear concentration profile given by
\beq
c(x,t=0)=
\begin{cases}
    c_I & \text{for}\quad 0\leq x\leq\lambda_1,\\
    c_I\frac{\lambda_2-x}{\lambda_2-\lambda_1} & \text{for}\quad \lambda_1\leq x\leq\lambda_2,\\
    0 & \text{for}  \quad \lambda_2 < x\leq 1,
\end{cases}
\label{eq:new1}
\eeq
Using \ref{eq:rie3} yields the initial velocity
\beq
u(x,t=0)=
\begin{cases}
    (c_I-\bar c)x & \text{for}\quad 0\leq x\leq\lambda_1,\\
    A_1x^2+B_1x+C_1 & \text{for}\quad \lambda_1\leq x\leq\lambda_2,\\
    \bar c(1-x) & \text{for}  \quad \lambda_2 < x\leq 1,
\end{cases}
\label{eq:new1}
\eeq
where $\bar c=c_I\frac{\lambda_1+\lambda_2}{2}$, and the constants are given by
\beq
    A_1=-\frac{c_I}{2(\lambda_2-\lambda_1)},\qquad B_1=\frac{c_I\lambda_2}{\lambda_2-\lambda_1}-\bar c,\qquad C_1=c_I\lambda_1 +\frac{c_I}{\lambda_2-\lambda_1}\left(\lambda_1\lambda_2+\lambda_1^2/2\right).
\eeq
Finding $\xi(x,t)$ from equation \ref{eq:chareq4} now gives, that
\beq
    \xi(x,t)=
    \begin{cases}
    \frac{x}{1+\frac 1\lambda(1-\lambda)(1-\exp(-\bar{c}t)} & \text{for} \quad x \in I_1,\\
    A_2\xi_2^2+B_2\xi_2^2+C_2 &  \text{for}\quad x\in I_2,\\
    \frac{x-1+\exp(-\bar{c}t)}{\exp(-\bar{c}t)} & \text{for} \quad x \in I_3,
    \end{cases}
\label{eq:new1}
\eeq
where
\beq
A_2=\frac{A_1}{\bar c}(1-\exp(-\bar c t)),\qquad B_2=1+\frac{B_1}{\bar c}(1-\exp(-\bar c t)),\qquad C_2=\frac{C_1}{\bar c}(1-\exp(-\bar c t)),
\eeq
and
\beq
\xi_2=\frac{-B_2+\sqrt{B_2^2-4A_2(C_2-x)}}{2A_2},
\eeq
where the plus solution has been chosen to ensure, that $\xi\to x$ as $t\to 0$.
Finally,
\begin{eqnarray}
    I_1&=&\left[0,\lambda_1+\frac {\lambda_1}{\bar c}(c_I-\bar c)(1-\exp(-\bar c t))\right],\\
    I_2&=&\left[\lambda_1+\frac {\lambda_1}{\bar c}(c_I-\bar c)(1-\exp(-\bar c t)),1-(\lambda_2-1)\exp(-\bar ct)\right],\\ I_3&=&\,\left[1-(\lambda_2-1)\exp\left(-\bar ct\right),1\right].
\end{eqnarray}
Plugging into \ref{eq:chareq3} gives $u(x,t)$ as
\beq
    u(x,t)=
    \begin{cases}
    \frac{(c_I-\bar c)\exp(-\bar c)x}{1+\frac 1\lambda(1-\lambda)(1-\exp(-\bar{c}t)} & \text{for} \quad x \in I_1,\\
    \left(A_1\xi_2^2+B_1\xi_2^2+C_1\right)\exp (-\bar c t) &  \text{for}\quad x\in I_2,\\
    \bar c\left(1-x\right) & \text{for} \quad x \in I_3,
    \end{cases}
\label{eq:new1}
\eeq
as shown in figure \ref{fig:analsolu1} along with $c$ found from equation \ref{eq:rie3}.

\subsection{Results for large M\"unch number} In the limit of large $M\gg 1$ we cannot neglect the pressure gradient along the channel and this term dominates the advective term in (\ref{eq:jfm1}), i.e. the second derivative in $u$. Thus \beq \partiel cx = -Mu \eeq \beq \partiel ct+\partiel{cu}{x}=\bar D \ppartiel c x \eeq giving the nonlinear diffusion equation \beq \partiel ct =M\partiel{}{x}\left[c\partiel cx\right] + D \ppartiel c x \eeq If we neglect molecular diffusion, which is true as long as $Mc \gg \bar D \approx 10^{-5}$,  the resulting universal nonlinear diffusion equation can be written \beq \partiel ct =M \partiel{}{x}\left[c\partiel cx\right] \label{eq:largeM} \eeq which belongs to a class of equations which have been studied e.g. in the context of intense thermal waves by Zeldovich et al. and flow through porous media by Barenblatt \cite[]{Barenblatt} in the 50'ies.
The M\"unch number $M$ can be removed by rescaling the time according to $T=Mt$, so in this limit we get very slow motion with a time scale growing linearly with $M$. The equation (\ref{eq:largeM}) admits scaling solutions of the form \beq
c(x,t) = (Mt)^{\alpha} \Phi (\xi)  \,\,\,\, {\rm with} \,\,\,\, \xi = x(Mt)^{\beta} \label{scaling-general} \eeq as long as $\alpha + 2 \beta + 1 =0$. The total amount of sugar is, however, conserved. In our rescaled units \beq
\int_0^1 c(x) \, {\rm d} x = \lambda
\label{sugar-cons}
\eeq
where, as before, $\lambda$ is the fraction of the tube initially containing the sugar.
We can only hope to find a scaling solution in the intermediate time-regime, where the precise initial condition has been forgotten, but the far end ($x=1$) is not yet felt. Thus we can replace the integral
(\ref{sugar-cons}) with
\beq
\int_0^\infty c \, {\rm d} x = \lambda
\label{conc-inf}
\eeq
which implies that $\alpha = \beta = -1/3$ and \beq
c(x,t) =  (Mt)^{-1/3} \Phi (\xi)  \,\,\,\, {\rm with} \,\,\,\, \xi = {\frac{x}{(Mt)^{1/3}}} \label{eq:scaling} \eeq Inserting this form into (\ref{eq:largeM}), we obtain the differential equation for $\Phi$ \beq \frac{1}{2} {\frac{d^2 \Phi^2}{d \xi^2}}+ \frac{1}{3} {\frac{d  (\xi
\Phi) }{d \xi}}=0
\label{eq:scale-dif}
\eeq
which can be integrated once to
\beq
\Phi {\frac{d  \Phi }{d \xi}}+ \frac{1}{3} \xi \Phi ={\rm const} \label{eq:scale-dif2} \eeq Due to the boundary condition $ \partial c / \partial x = 0$  in the origin, the constant has to vanish and we find the solution \beq
\Phi(\xi) = {\frac{1}{6}}(b^2 - \xi^2)
\label{eq:scale-sol}
\eeq
which is valid only for $\xi$ less that the constant $b$. For $\xi > b$, $\Phi$ is identically 0. The fact that the solution - in contrast to the linear diffusion equation - has {\it compact support}, is an interesting characteristic of a large class of nonlinear diffusion equations \cite[]{Barenblatt}. The value of $b$ is determined by the conservation integral (\ref{conc-inf}) giving  $\int_0^\infty \Phi \, {\rm d} \xi = 1$, and thus $b= (9 \lambda)^{1/3} $.

\begin{figure}
\centering
\centerline{\includegraphics[width=1\textwidth]{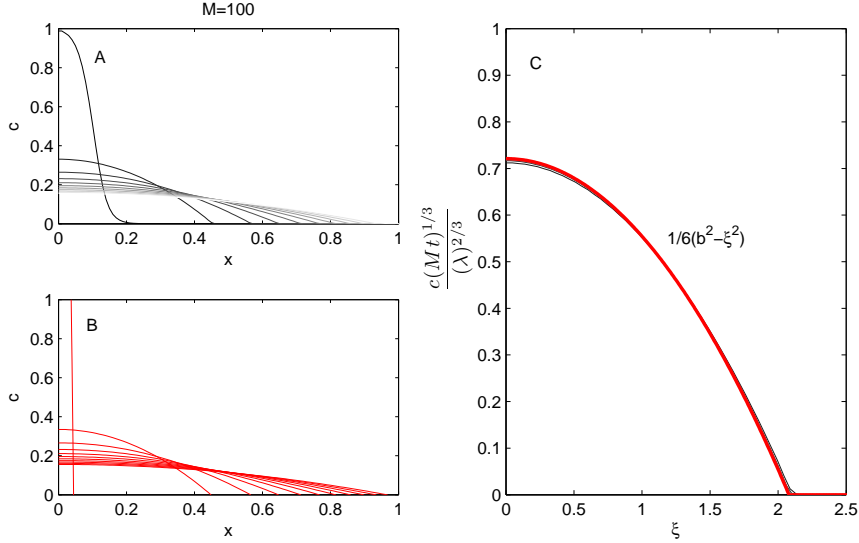}}
\caption{(A) Numerical simulation of equation (\ref{eq:largeM}) compared with (B) the scaling solution (\ref{eq:scale-sol-fin}) and (C) (\ref{eq:scale-sol}). The intial condition has the form $c(x,0)=1-\frac{1}{1+\exp\left(-\frac{x-\lambda}{\epsilon}\right)}$
where $\lambda=0.1$ and $\epsilon=2\times 10^{-2}$ and the curves are equidistant in time. When $\lambda$ controlling the size of the region of nonzero initial sugar concentration becomes larger, a more accurate scaling solution is found by letting $t \rightarrow t+t_0$ and treating $t_0$ as an unknown parameter. In (C), we have omitted the first curve (the initial condition).}
\label{fig:largeM}
\end{figure}

The final solution thus has the form
\beq
    c(x,t) =
    \begin{cases}
       {\frac{1}{6 M t}} \left( (x_f(t))^2  - x^2  \right) & \textrm{for}\quad x< x_f(t) = (9 \lambda Mt)^{1/3}  \\
       0 & \textrm{for}\quad x> x_f(t)
    \end{cases}
\label{eq:scale-sol-fin}
\eeq
which shows that the sugar front moves as $x_f (t) \sim t^{1/3}$ and the concentration at the origin decays as $c(0,t) \sim t^{-1/3}$. To check the validity of this solution, also when the initial condition has support in a finite region near the origin, we plot $({\bar c} L)^{-2/3} (Mt)^{1/3} c(x,t)$ against $\xi = x({\bar c} L Mt)^{-1/3}$ in figure \ref{fig:largeM}, C.

\subsection{Numerical methods for nonzero $M$ and $\bar D$}
\begin{figure}
\centering
\centerline{\includegraphics[width=1\textwidth]{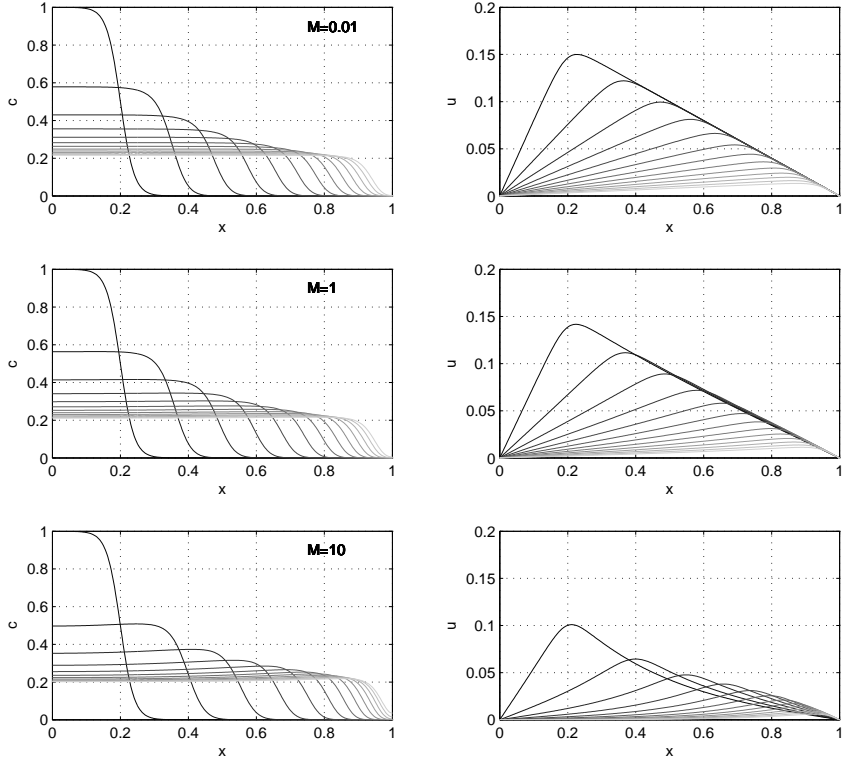}}
\label{fig:numsim1}
\caption{Results of numerical simulation of equation \ref{eq:n4} using the boundary conditions $u(0,t)=u(1,t)=0$ for different values of $M$. $\bar D$ is kept constant at $10^{-5}$. The initial condition was $c(x,0)=1-\frac{1}{1+\exp\left(-\frac{x-\lambda}{\epsilon}\right)}$ where $\lambda=0.2$ and $\epsilon=2\times 10^2$}
\end{figure}

For nonzero values of $M$ and $\bar D$, the equations of motion,
\beq
    \ppartiel ux -Mu=\partiel cx
\label{eq:n1}
\eeq
and
\beq
    \partiel ct +\partiel {cu}{x}=\bar D\ppartiel cx
\label{eq:n2}
\eeq
cannot be solved analytically. However, they can be written as a single integro-differential equation, which is straightforward to solve on a computer. If we choose a set of linear boundary conditions, $B_x[u]=a_i$, for equation \ref{eq:n1}, the solution can be written as
\beq
    u=\int_0^1G(x,\xi)\partiel c\xi d\xi + u_2.
\label{eq:n3}
\eeq
Here, $G(x,\xi)$ is the Green's function for the differential operator $\ppartiel{}x-M$ with boundary conditions $B_x[u]=0$ and $u_2$ fulfills the homogeneous version of \ref{eq:n1} with $B_x[u]=a_i$. Plugging this into equation \ref{eq:n2} yields
\beq
    \partiel ct +\partiel{}{x}\left(c\left(\int_0^1G(x,\xi)\partiel c\xi d\xi + u_2\right)\right)=\bar D \ppartiel cx
\label{eq:n4}
\eeq
For the closed tube, ie. for the boundary conditions $u(0,t)=u(1,t)=0$, $G(x,\xi)$ is given by
\beq
    G(x,\xi)=
    \begin{cases}
        -\frac{\sinh(a(1-x))}{a\sinh a}\sinh a\xi & \textrm{for}\quad\xi<x, \\
        -\frac{\sinh ax}{a\sinh a}\sinh (a(1-\xi)) & \textrm{for}\quad \xi>x,
    \end{cases}
\label{eq:n5}
\eeq
and $u_2=0$. To increase numerical accuracy, it is convenient to transform equation \ref{eq:n4} by defining
\beq
    \frac{\partial f}{\partial x}=c-\bar c
\label{eq:n6}
\eeq
 and choosing $f(0)=f(1)=0$ such that $f(x) =\int_0^x (c-\bar c)\ud \xi$. Inserting in equation
\ref{eq:n4}, we get that
\beq
    \frac{\partial f}{\partial t} =
    \bar{D}\ppartiel fx-\left(f(x)-\int_0^1
    \frac{\partial K(x,\xi)}{\partial{\xi}}f(\xi)\ud \xi\right)\left(\frac{\partial f}{\partial x}+\bar c\right),
\label{eq:n7}
\eeq
where
\beq
    \frac{\partial K(x,\xi)}{\partial \xi}=
    \begin{cases}
    -a\frac{\sinh(a(1-x))}{\sinh a}\sinh a\xi & \textrm{for}\quad\xi<x, \\
    -a\frac{\sinh ax}{\sinh a}\sinh (a(1-\xi)) & \textrm{for}\quad \xi>x.
    \end{cases}
\label{eq:n8}
\eeq
To solve equation \ref{eq:n7} we used \textsc{Matlab}'s built-in time solver \texttt{ode23t} which is based on an explicit Runge-Kutta formula along with standard second order schemes for the first and second order derivatives. For the spatial integration, the trapezoidal rule was used \cite[]{Press:2001}. The numerical code can be found in appendix \ref{app:num}. Results of a numerical simulation for different values of $M$ is shown in figure \ref{fig:numsim1}
\section{Comparison between theory and experiment}
In section \ref{sec:results}, we have presented experiments demonstrating the movement of a sugar solution inside a membrane tube surrounded by a reservoir of water. We now wish to consider whether the theory is in agreement with the experimental results.
\subsection{Setup I}
 \begin{figure}
\centerline{\includegraphics[width=1\textwidth]{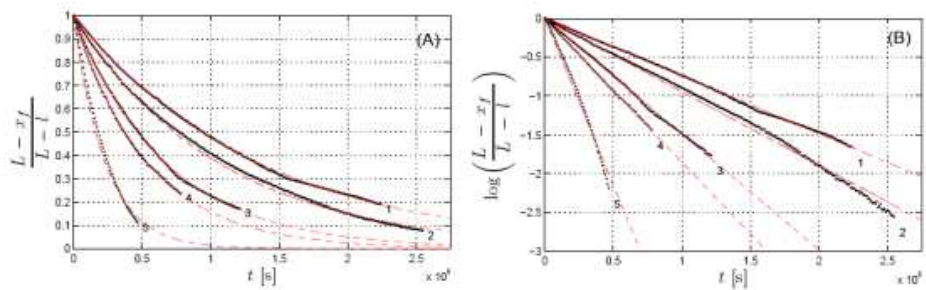}}
\caption{(A): Experimental (black dots) and fits to equation \ref{eq:chareq13} for the relative front position vs. time. (B): Semi-logarithmic version of (A).}
\label{fig:setupIdatavsmodel}
\label{fig:muncht0}
\end{figure}
 \begin{figure}
\centerline{\includegraphics[width=0.75\textwidth]{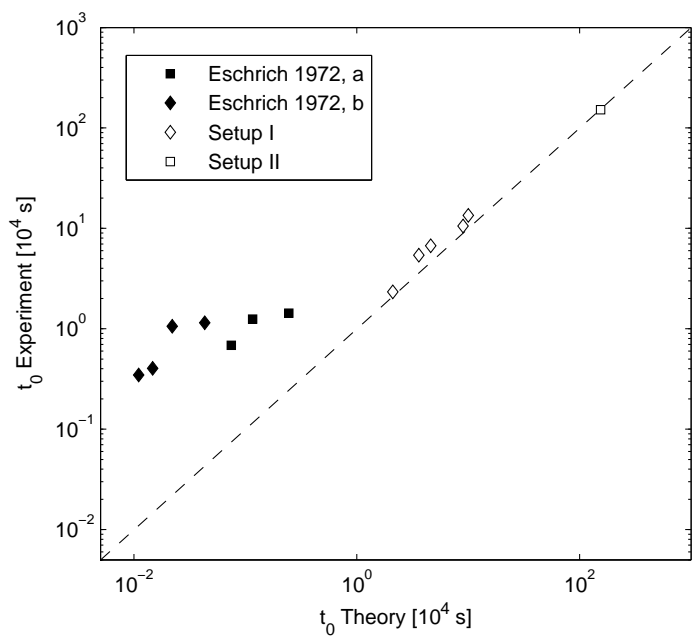}}
\caption{Our experimentally obtained values of $t_0$ plotted together with the results found by Eschrich \etal \cite[]{Eschrich:1972}. Data points marked with an a represents results from closed tube experiments and points marked with a b represents results from semi-closed experiments take from the original paper, figures 8 and 9.}
\label{fig:muncht0}
\end{figure}
The plot in figure \ref{fig:setupIdatavsmodel} shows the relative front position, $\frac{L-x_f}{L-l}$,
plotted against time for five different experiments conducted with setup I. The numbers $1-5$ indicates the sugar concentrations used, cf. table \ref{table:setupIresults}. One clearly sees, that the relative front position approaches zero faster for high concentrations than for low. Typical values of $M$ and $\bar D$ are $M\sim 10^{-8}$ and $\bar D\sim 10^{-5}$, so it is reasonable to assume that we are in the domain where the analytical solution for $M=\bar D=0$ is valid.
To test the result from equation \ref{eq:chareq10} against the experimental data, the plot in figure \ref{fig:setupIdatavsmodel} shows the logarithm of the relative front position plotted against time.
For long stretches of time the curves are seen to approximately follow straight lines in good, qualitative agreement with theory.
The red dashed lines are fits to equation (\ref{eq:chareq10}), and we interpret the slopes as $-\frac 1{t_0}$, the different values plotted in figure \ref{fig:muncht0} against the theoretical values.
The theoretically and experimentally obtained values of $t_0$ are in good quantitative agreement, within 10-30\%. Generally, theory predicts somewhat smaller values of $t_0$ than observed, implying that the observed motion of the sugar front is a little slower than expected from the pressure-flow hypothesis. Nevertheless, as can be seen in figure \ref{fig:muncht0} these results are a considerable improvement to the previous results obtained by Eschrich \etal as we find much better agreement between experiment and theory.
\subsection{Setup II}
\begin{figure}
\centerline{\includegraphics[width=1\textwidth]{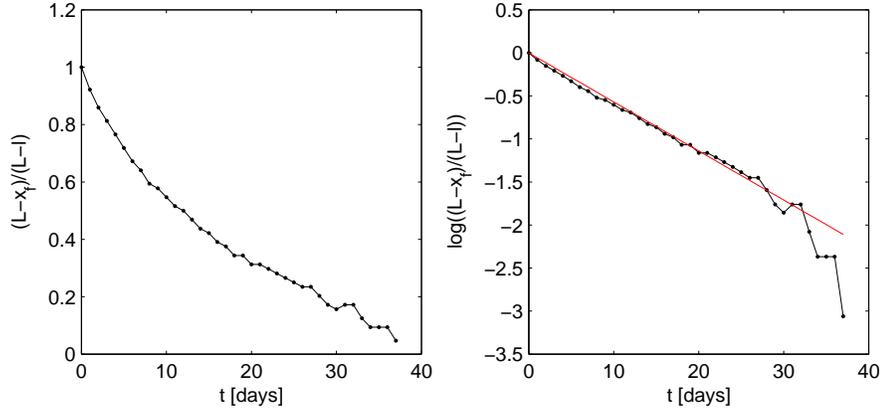}}
\caption{Results from setup II showing the relative front position as a function of time.}
\label{fig:setupIIdataproc1}
\end{figure}

The plot in figure \ref{fig:setupIIdataproc1} shows the relative front position, $\frac{L-x_f}{L-l}$,
plotted against time for the experiment conducted with setup I. On the semi-logarithmic plot, the curves are seen to follow straight lines in good, qualitative agreement with the simple theory for $M=\bar D=0$. As can be seen in figure \ref{fig:muncht0}, we also found very good quantitative agreement between the experiment and theory for setup II.

\begin{figure}
\centerline{
\includegraphics[width=0.5\textwidth]{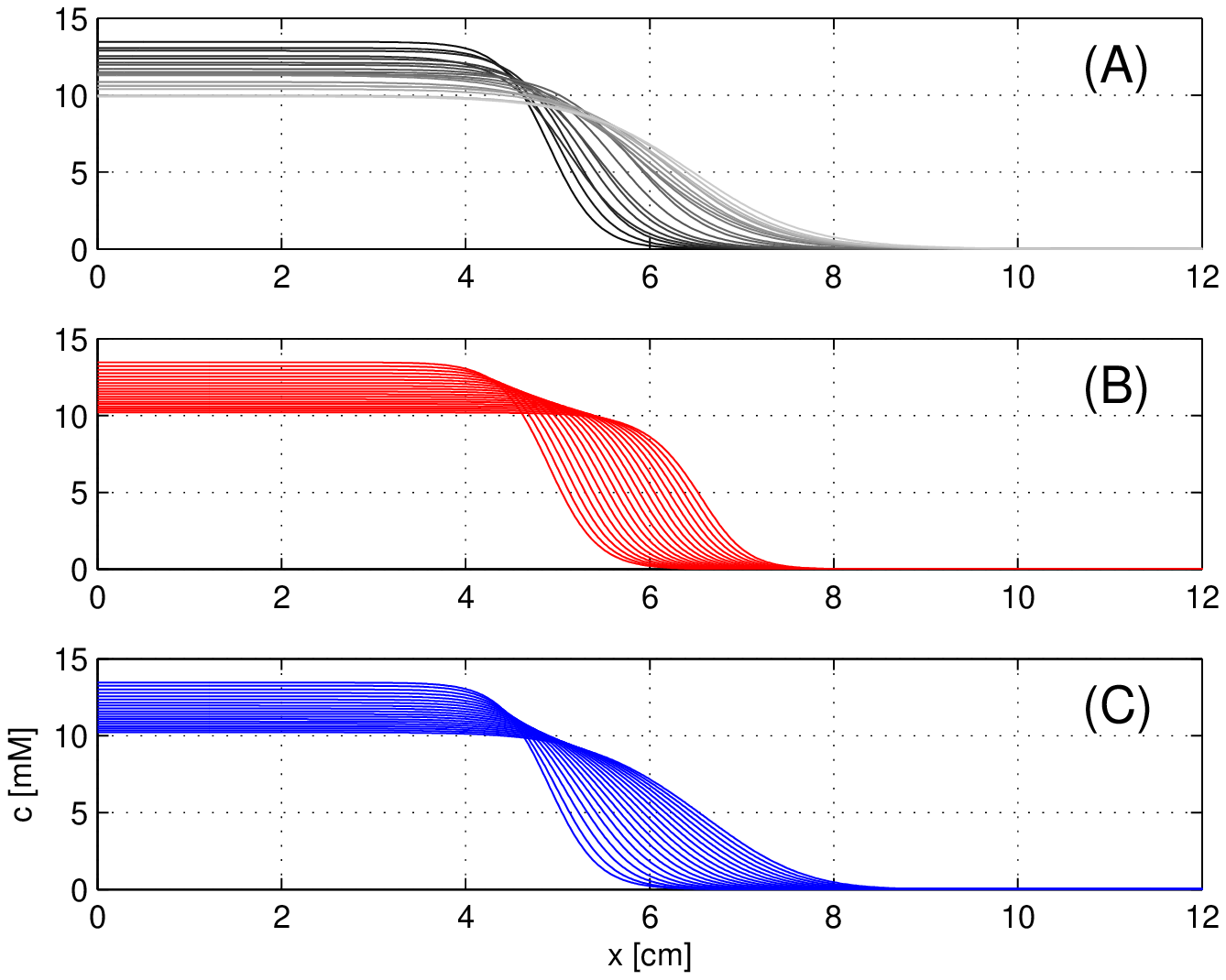}
\includegraphics[width=0.5\textwidth]{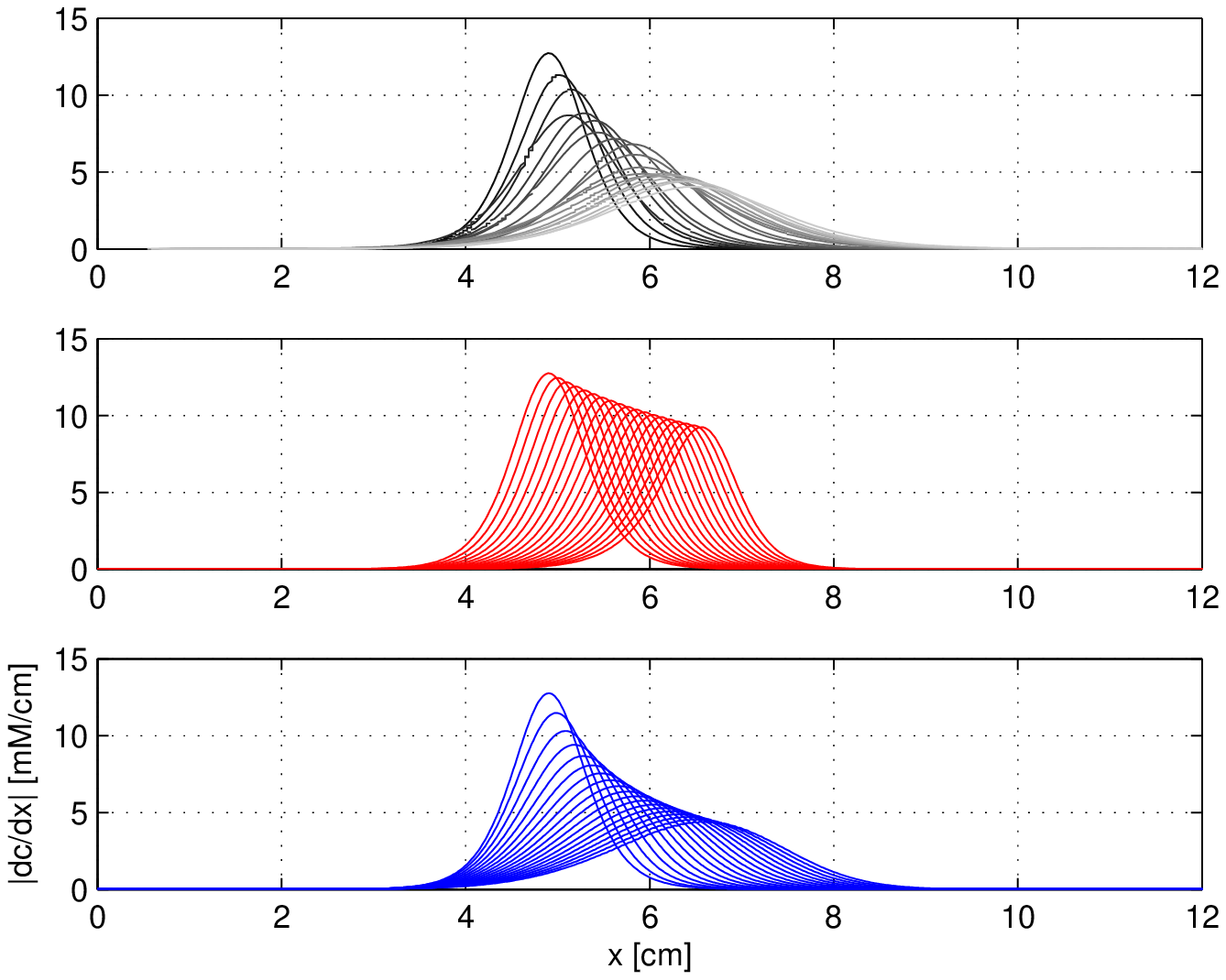}}
\caption{Results from setup II showing the experimental data (top row) and the numerical model for $M=D=0$ (middle row) and for $M=10^{-9},\; D=6.9\times 10^{-11}\text{m}^2\text{s}^{-1}$ (bottom row).}
\label{fig:setupIIdataproc2}
\end{figure}

To test how well the motion of the sugar front observed in the experiments with setup II was reproduced by our model, we solved the equations of motion numerically starting with the initial conditions from figure \ref{fig:riocomp1}. For $M=\bar D=0$, The results are shown as (red) curves in figure \ref{fig:setupIIdataproc2} (B). While the front positions are reproduced relatively well, the shape of the front is not, so diffusion must play a role. This can be seen in figure \ref{fig:setupIIdataproc2} (C) which shows the result of a simulation with $M=10^{-9},\; D=6.9\times 10^{-11}\text{m}^2\text{s}^{-1}$. Clearly, the model which includes diffusion reproduces the experimental data significantly better.

To study the shape of the front in greater detail, consider the plots on the right in figure \ref{fig:setupIIdataproc2}. Here the gradient of the concentration curves on the left in figure \ref{fig:setupIIdataproc2} are shown. In (A) we clearly see a peak moving from left to right while it gradually broadens and flattens. In (B) we also see the peak advancing, but the flattening and broadening is much less pronounced. In (C) we see that the model which includes diffusion reproduces the gradual broadening and flattening of the front very well.

\section{Summary}
In this paper we have studied osmotically driven, transient pipe flows. The flows are generated by concentration differences of sugars in closed tubes, fully or partly enclosed by semi-permeable membranes surrounded by pure water.  The flows are initiated by a large concentration in one end of the tube and we study the approach to equilibrium, where the sugar is distributed evenly within the tube. Experimentally, we have used two configurations: the first is an updated version of the setup of Eschrich \etal  where the flow takes place in a dialysis-tube and the sugar is followed by introducing a dye. The advantage is the relatively rapid motion, due to the large surface area. The disadvantage is that the sugar concentration cannot be inferred accurately by this method and for this reason we have introduced our second setup, where the sugar concentration can be followed directly by refraction measurements. On the theoretical side, we first re-derive the governing flow equations and introduce the dimensionless M\"unch number $M$. We then show that analytical solutions  can be obtained in the two important limits of very large and very small $M$. In the general case we show how numerical methods based on Green's functions are very effective. Finally, we compare theory and experiment with
very good agreement. In particular the results or the velocity of the front (as proposed by Eschrich et al.) can be verified rather accurately.
\begin{acknowledgments}
It is a pleasure to thank Francois Charru, Marie-Alice Goudeau-Boudeville, Herv\'e Cochard, Pierre Cruiziat, Alexander Schulz, N. Michelle Hollbrook and Vakhtang Putkaradze for many useful discussions. Much appreciated technical assistance was provided by Erik Hansen.
\end{acknowledgments}
\appendix
\section{Setup and Methods}
\label{app:setups}
\subsection{Materials}
\subsubsection{Chemicals}
The sugar used was a dextran (Sigma-Aldrigde, type D4624) with an average molecular weight of 17.5 kDa. The dye used was a red fruit dye (Flachsmann Scandinavia, R\o d Frugtfarve, type 123000) consisting of an aquous mixture of the food additives E-124 and E-131 with molecular weights of 539 Da and 1159 Da respectively \cite[]{Pubmed:2007}. Even though the molecular weights are below the MWCO of the membrane, the red dye were not observed to leak through the membrane. This however, was observed when using another type of dye, Methylene blue, which has a mulecular weight of 320 Da.
\subsubsection{Membrane}
The membrane used in setup I was a semipermeable dialysis membrane tube (Spectra/Por Biotech cellulose ester dialysis membrane) with a radius of 5 mm, thickness 60 $\mu$m and a MWCO of 3.5 kDa. The membrane used in setup II had identical specifications except that it had a radius of 3 cm allowing it to cover the interface between the prism and the water reservoir after beeing cut in half.
\subsection{Elastic and osmotic properties of the materials used}
\label{app:vy}
\subsubsection{Elastic properties of the membrane tube}
Figure \ref{fig:osmoticstrength} (right) shows the relation between internal pressure and radius for the membrane tube. For pressures less than $1.2$ bar, a linear relation between the relative radial increase and internal pressure was found. Linear elasticity theory \cite[]{Love:1944} predicts that for a thin-walled cylindrical tube
\beq
    r=r_0+\frac{r_0^2p}{dE},
\eeq
where $r_0$ is the equilibrium radius and $d$ the thickness of the tube, $p$ is pressure and $E$ is Young's modulus of the membrane tube.
From this, $E$, was found to be
\beq
    E=0.66\pm 0.01 \text{ GPa}
\eeq
%
\subsubsection{Osmotic strength of Dextran}
\begin{figure}
\centerline{\includegraphics[width=0.45\textwidth]{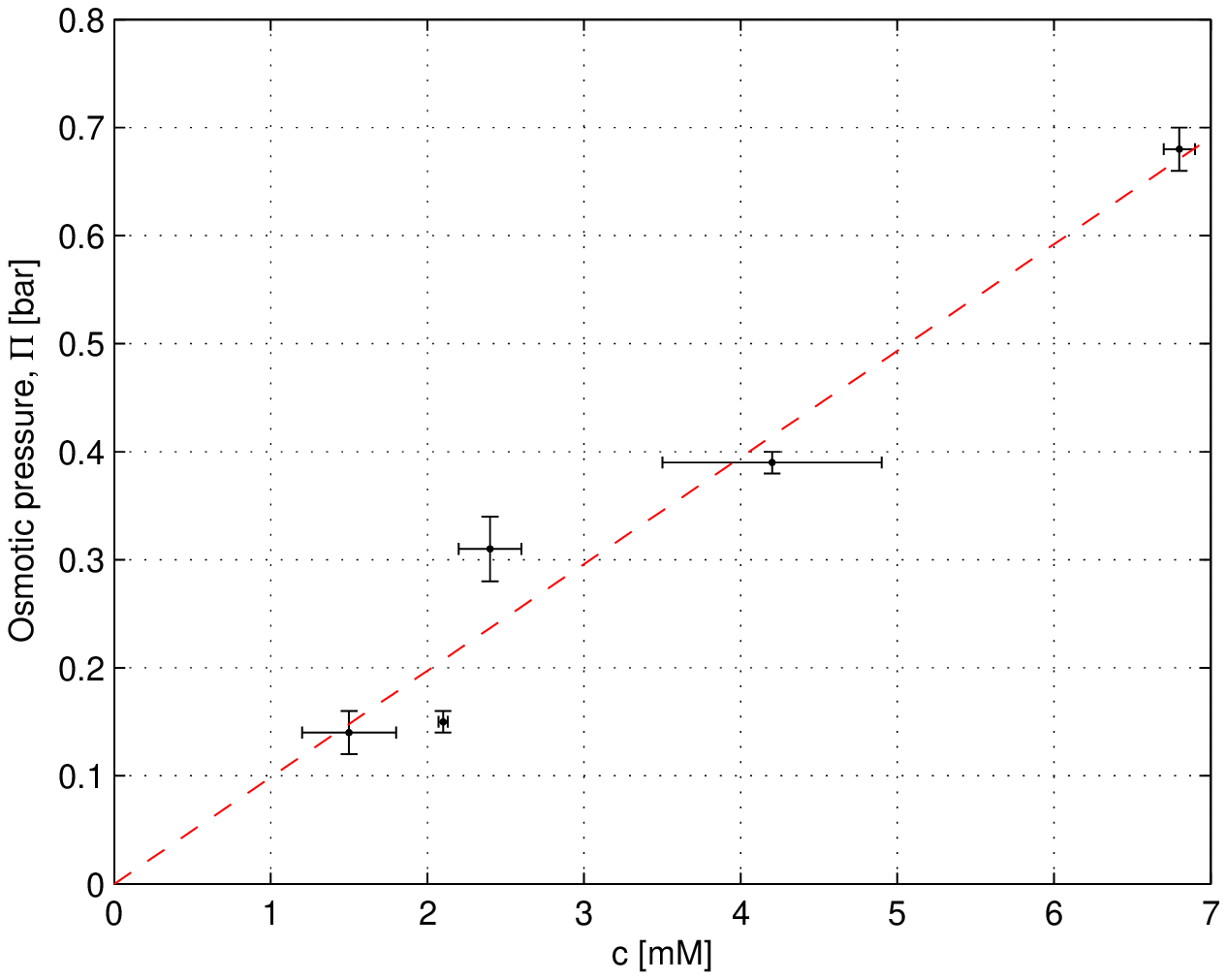}\includegraphics[width=0.45\textwidth]{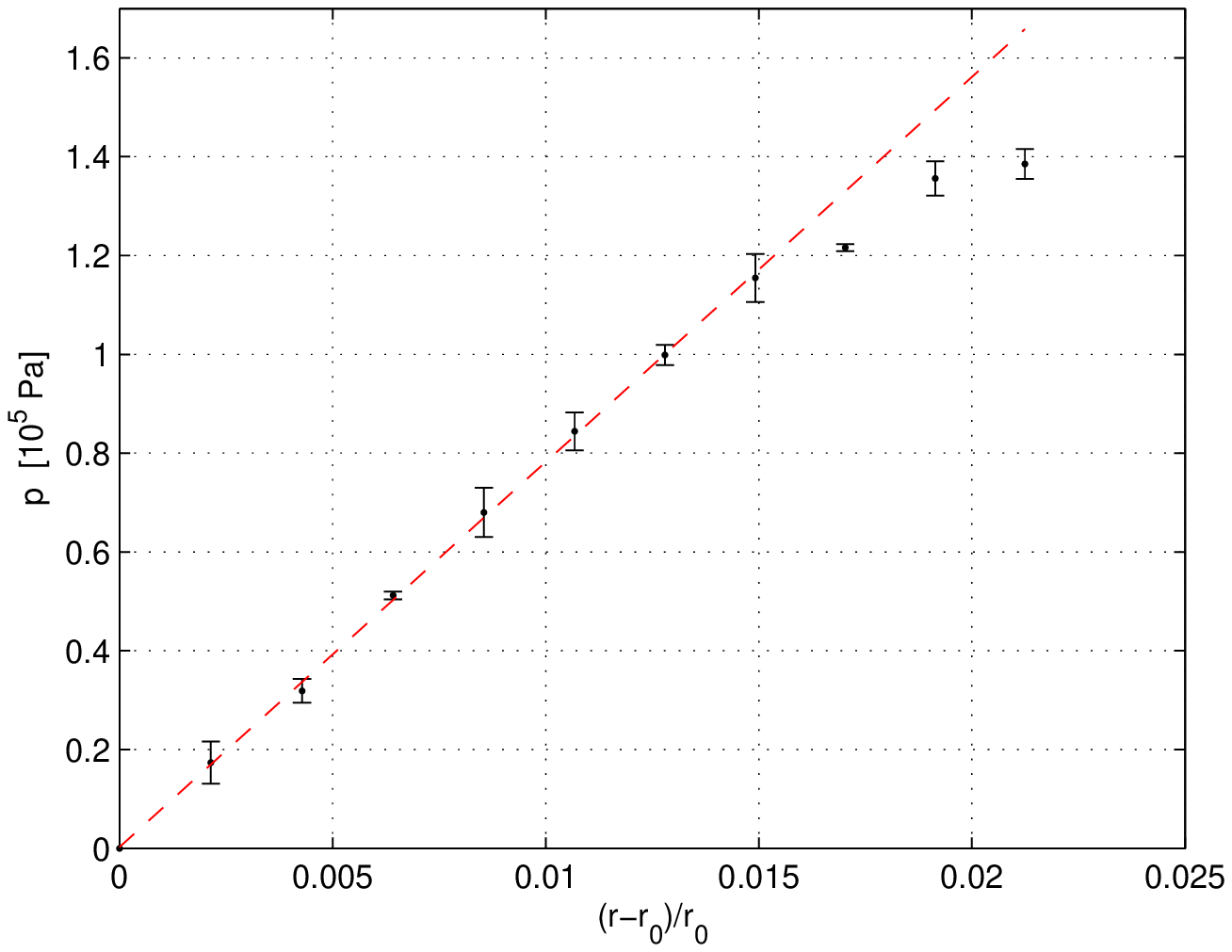}}
\caption{Left: van't Hoff relation for $^{17.5 \text{kDa}}$Dextran. Right: Elastic properties of the membrane tube}
\label{fig:osmoticstrength}
\end{figure}

Figure \ref{fig:osmoticstrength} (left) shows the relation between dextran concentration and osmotic pressure found from the experiments shown in figure \ref{fig:setupIresults}. A linear fit gives
\beq
    \Pi = (0.1\pm 0.01 \text{ bar mM}^{-1})c
\eeq
where $\Pi$ has units of bar, and $c$ is measured in mM. This is in good agreement with values given by \cite[]{Jonsson:1986}

\section{Tracking the sugar front position}
\subsection{Setup I}
After running an experiment with setup I, the raw data we had acquired consisted of a series of pictures as shown in figure \ref{fig:setupIresults}.

\begin{figure}
\centering
\centerline{\includegraphics[width=1\textwidth]{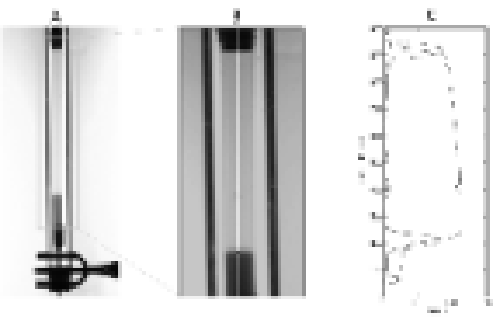}}
\caption{Data processesing in setup I.
\newline(A): Raw RGB data image showing the sugar solution mixed with red dye inside the membrane tube.
\newline (B): The image from (A) has been cropped and filtered.
\newline (C): The solid black curve is the intensity of color in (B) taken along the white vertical line. The red curve is the absolute value of the gradient of the intensity. The front position is taken to be where this curve has its maximum value (at approximately 200 pixels). The black dot in (B) is the position of the front found in this manner. The peak in the intensity gradient near 920 pixels is due to the membrane fitting, and was ignored.}\label{fig:dataproc}
\end{figure}

To track the position of the sugar-dye front, the images was treated as shown in figure \ref{fig:dataproc}.
First, the image was imported into \textsc{Matlab} as (A), and then cropped to shown only the membrane tube, (B). Simultaneously, it was filtered to give the highest contrast for obtaining a well-defined front position. To find the front position, a vertical line running along the center of the membrane tube was picked out, shown as a white line in (B). Along this line, the color intensity was found, shown as the black curve in (C). Finally, the gradient of the color intensity was found -- shown in (C) as the red curve -- and the front position was defined to be the the position of the maximum in intensity gradient.

\begin{figure}
\centerline{\includegraphics[width=1\textwidth]{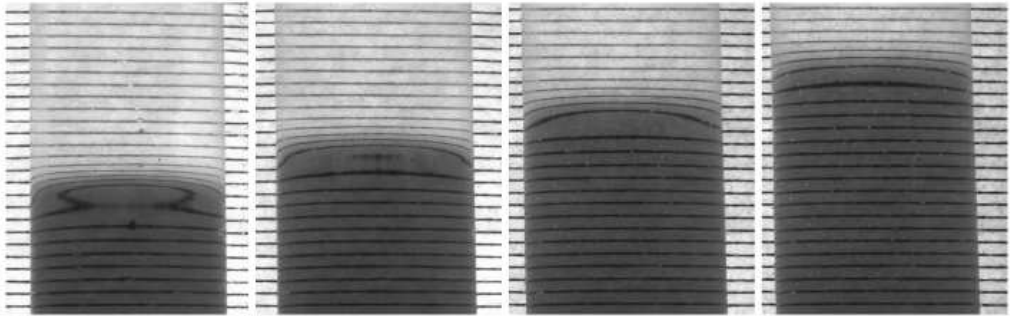}}
\caption{Closeup images of the sugar front as it moves. It is clearly seen, that the dye moves with the point at which the concentration gradient is largest.}
\label{fig:sugartracking}
\end{figure}

\label{app:setupI}
To justify the use of red dye as the tracking medium, we took closeup images of the sugar front as shown in figure \ref{fig:sugartracking}. It is clearly seen, that the dye moves with the point at which the concentration gradient is largest. Thus, we conclude that the dye travels along with the sugar.
\subsection{Setup II}
\label{app:setupII}
A camera recorded images of the screen at regular intervals, as shown in figure \ref{fig:expres2}. The deflection at the bottom of the image corresponds to a high sugar concentration inside the lower part of the prism, and the vertical deflection is due to a strong gradient in index of refraction near the sugar front.
\begin{figure}
\centerline{\includegraphics[width=1\textwidth]{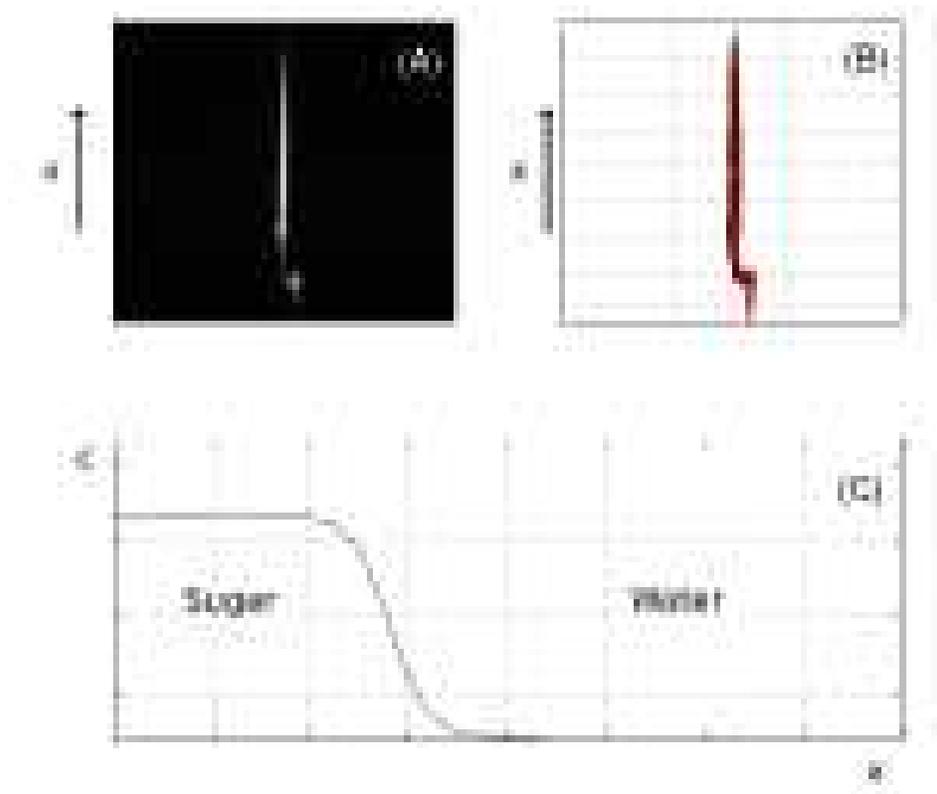}}
\caption{Data processing in Setup II.
(A): Raw data image.
(B): Filtered data image.
(C): Index of refraction inside the prism giving the red curve in (B) cf. equation \ref{eq:delta1} and \ref{eq:delta2} }
\label{fig:rioraw}
\end{figure}

\begin{figure}
\centering
\includegraphics[width=1\textwidth]{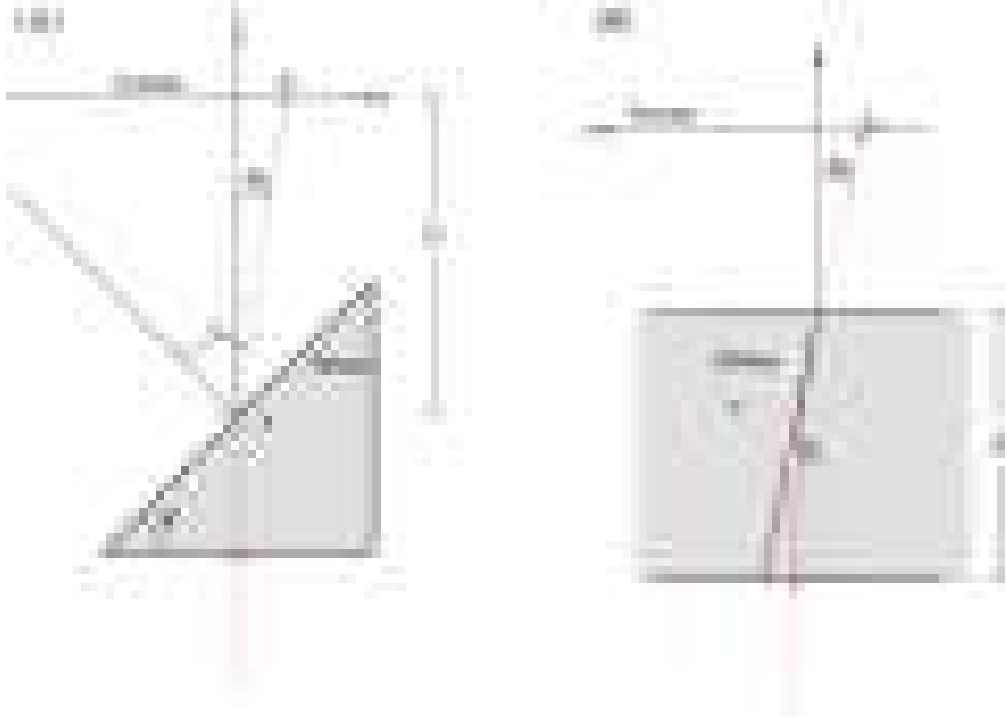}
\caption{Deflection of the laser beam as i passes through the prism.}.\label{fig:setupIIgeom}
\end{figure}

\label{sec:trackfrontII}
As the beam passes through the prism, it gets deflected due to variations in index of refraction of the fluid inside the prism relative to the surrounding air.
To determine the horizontal deflection of a light beam passing through the prism, we consider the situation sketched in figure \ref{fig:setupIIgeom}. The deflection $\Delta_1$ along the $y$ axis is given by
\beq
    \Delta_1=G\tan\Theta_1
\eeq
where $G$ is the orthogonal distance from the prism to the screen, and $\Theta_1$ is the deflection angle as shown in the figure. To find $\Theta_1$ we notice that
\beq
    r_1-\Theta_1=\Psi,
\label{eq:theta1}
\eeq
where $\Psi$ is the prism angle. The deflection angle, $r_1$, is given by the Snell-Descartes law \cite[]{Hect}
\beq
    n\sin= \Psi\sin r_1,
\label{eq:snell}
\eeq
where $n$ is the index of refraction of the fluid inside the prism. In the experiments, $\Theta_1$ is typically small so
\beq
n \sin \Psi=\sin r_1=\sin (\Psi+\Theta_1)\simeq\sin \Psi+ \Theta_1\cos \Psi.
\eeq
so the horizontal deflection
\beq
\Delta_1\simeq G\Theta_1=G\left(n-1\right)\tan \Psi.
\label{eq:delta1}
\eeq

To determine the vertical deflection consider the situation sketched in figure \ref{fig:setupIIgeom}, (B). According to Fermat's principle, light travels along the path that can be traversed in the least possible time. Consequently, as the light passes through the prism, it travels in a circular arc with a radius of curvature, $\mathcal R$, given by
\beq
    \frac 1{\mathcal R}=\mathbf N \cdot \frac{\nabla n}{n},
\label{eq:fermat}
\eeq
where $\mathbf{N}$ is the normal to the trajectory of the light beam \cite{Landau:1984}.
The vertical deflection is given by
\beq
    \Delta_2=G\tan \Theta_2
\eeq
where the angle $\Theta_2$ is given by
\beq
    \sin\Theta_2=n\sin r_2
\eeq
Since $r_2$ is small, it is clear from the figure that $\sin r_2\simeq \frac{g}{\mathcal R}$, so from equation (\ref{eq:fermat})
\beq
    \sin r_2 \simeq \frac g{\mathcal R} = \frac g n \partiel nx.
\eeq
In the limit where $r_2$ and $\Theta_2$ are small we get for the vertical deflection
\beq
    \Delta_2=Gg\partiel nx,
\label{eq:delta2}
\eeq
where we have assumed that $\nabla n$ is constant as the beam traverses the prism.
Having obtained $\Delta_1$ and $\Delta_2$ we now have to deduce $n(x)$ from the projected image. We will do this by assuming that $n(x)$ has a generic, sigmoid shape
\beq
    n(x)=n_0+n_1\left(1-\frac{1}{1+\exp{\left(-\frac{x-l}{\epsilon}\right)}}\right)
\label{eq:n}
\eeq
where the constants $n_0$ and $n_1$ controls the magnitude of $n$ and $l$ and $\epsilon$ controls the position and steepness of the front. A plot of this function can be seen in figure (\ref{fig:rioraw}, C)

The procedure for obtaining $n(x)$ was as follows. First, the raw data image was loaded into \textsc{Matlab} as shown in figure \ref{fig:rioraw}, (A). Then, the image was filtered to show only regions of high light intensity, shown as the black dots in (B). Then, a guess of the the form \ref{eq:n} was made, and the deflections $\Delta_1$ and $\Delta_2$ was calculated from equations \ref{eq:delta1} and \ref{eq:delta2}, shown as the red curve in (B). Finally, an optimization of the parameters was made using \textsc{Matlab}'s \texttt{fminsearch} engine, thereby giving the $n(x)$ of the form \ref{eq:n} best able to reproduce the image seen in (A). Generally, the assumption that $n$ was of the form \ref{eq:n} gave very good fits, as can be seen in figure \ref{fig:rioraw}, (B).

\section{Generalization of the equations of motion to non-cylindrical geometries}
\label{sec:othergeom}
When deriving equations \ref{eq:volumeconservationnondim2}-\ref{eq:soluteconservationnondim2} we have assumed that our system consisted of a cylindrical tube with semipermeable walls. The assumption of a cylindrical tube, however, have only been used in equation (\ref{eq:floweq8})  where we assumed that the cross-section area to perimeter ratio was
\beq
    \frac AS=\frac r2
\eeq
and in equation (\ref{eq:floweq9}) where we assumed the axial resistance in Stokes flow to be inversely proportional to the cross-section area
\beq
    \frac{2\pi}A=\frac  8{r^2}
\eeq
These two factors are of purely geometrical nature and appear in $M$, $\bar D$ and $t_0$ as
\beq
    M\propto  \frac 2r\cdot\frac{8}{r^2}=\frac{16}{r^3},
\eeq
\beq
    \bar D\propto \frac r2
\eeq
and
\beq
    t_0 \propto \frac r2.
\eeq
Therefore, as long as the assumption of a 1D flow velocity and concentration holds inside the tube the equations of motion can be extended to include other geometries, e.g. triangular tubes as used in setup II, by replacing the geometric factors in $M$, $\bar D$ and $t_0$ as discussed above.

Finding the cross-section area to perimeter ratio is trivial, and the expression for the axial resistance in  Stokes flow generally has the form
\beq
    u=\frac 1\eta\frac{A}{\alpha}\partiel px
\eeq
where $\frac{A}{\alpha}$ is a purely geometric factor, which for a cylindrical tube is $\frac {\pi r^2}{8\pi}$.
For various pipe cross-sections Mortensen \etal \cite[]{Mortensen:2005} has found $\alpha$ as a function of the dimensionless compactness
\beq
    C=\frac{S^2}A.
\eeq
\subsection{Setup II}
To extend the equations of motion to setup II, consider the following. For a isosceles right triangle with two sides of length $s$ and one of length $\sqrt 2 s$, we get that
\beq
    C=\frac{2(2s+\sqrt 2s)^2}{s^2}=12+8\sqrt 2
\eeq
Mortensen et. al. showed that for pipes with triangular cross-sections
\beq
    \alpha =\frac{25}{17}C+\frac{40\sqrt 3}{17}
\eeq
so in our case
\beq
    \alpha = \frac{300}{17}+\frac{200\sqrt 2}{17}+\frac{40\sqrt 3}{17}\simeq 38.36.
\eeq
Also,
\beq
    \frac AS=\frac{s}{4+2\sqrt 2}.
\eeq
Plugging into the expressions for $M,\; \bar D$ and $t_0$ we get
\beq
    M^{II}=\frac{38.36(8+4\sqrt 2)\eta  L^2L_p}{(2+\sqrt 2)s^3}=153.44\frac{\eta L^2L_p}{s^3},
\eeq
\beq
    \bar D^{II} =\frac{(2+\sqrt 2)Ds}{(4+2\sqrt 2)2RTc_0L^2L_p}=\frac{Ds}{2RTc_0L^2L_p}
\eeq
and
\beq
    t_0^{II}=\frac{(2+\sqrt 2)s}{(4+2\sqrt 2)L_pRTc_0 }=\frac{s}{2L_pRTc_0 }.
\eeq
The extra factors of $2+\sqrt 2$ comes from the fact that the membrane only covers one wall of length $a$, thereby scaling $L_p$ down to $\frac{L_p}{2+\sqrt 2}$.
\section{Numerical code}
\label{app:num}
\begin{verbatim}
%%%% 25 Feb 2008 %%%%
%%%% Numerical code for solving the equations of motion %%%%
%%%% for osmotically driven flows in a closed tube %%%%
%%%% Download at http://www.fysik.dtu.dk/~tbohr/munchsolver.m %%%%

function numsolver

%%%% Choose M and D %%%%
D = 1e-8;
M = 1e-8;
a=sqrt(M) ;

%%%% Initialize x and t %%%%
Nx = 500; %
x = linspace(0,1,Nx) ;
Nt = 100;
tspan = linspace(0,40,Nt ) ;

%%%% Initial condition %%%%
epsilon=2e-2;
x0=0.2;
c0=1-1./(1+exp(-(x-x0)/epsilon ));
cbar=trapz (x,c0) ;
for i=2:Nx
    f0(i) = trapz(x(1:i),c0(1:i).cbar);
end

%%%% Run solver %%%%
[t,f] = ode23t (@fderiv,tspan,f0,[],D,a,cbar,x,Nx) ;

%%%% Plot results %%%%
%% Calculate c %%
for i=1:Nt
    c(i,:)=gradient(f(i,:),x)+cbar;
end
%% Plot c %%
figure(1)
surf(c)
\end{verbatim}
\begin{verbatim}

%%%% Function fderiv for use in ode23t %%%%
function df = fderiv(t,f,D,a,cbar,x,Nx);
    P = zeros(Nx,1);
    f(1) = 0;
    f(Nx) = 0;
    temp = f';
    f = temp;
    %% Calculate P %%
    for i=2:Nx-2
        P(i) = f(i).(a/sinh(a)).(trapz(x(1:i),...
        sinh(a.(1.x(i)))..f(1:i).....
        sinh(a..x(1:i)))+trapz(x(i+1:end),...
        sinh(a..(1.x(i+1:end)))..f(i+1:end).....
        sinh(a.x(i))));
    end
    %% Impose boundary conditions on P %%
    P(1) = 0;
    P(Nx) = 0;
    P(Nx-1) = P(Nx-2)/2;
    P=P';
    laplacef = 4.del2(f,x);
    laplacef(1) = 0;
    laplacef(Nx) = 0;
    %% Update f %%
    df=D*laplacef-P*.(cbar+gradient(f,x));
    %% Impose boundary conditions on f %%
    df(1) = 0;
    df(Nx) = 0;
\end{verbatim}

\section{Front propagation in an elastic tube}
\label{sec:elastica}
In this section we shall investigate the effect of expansion of the tube. First we simply assume that water can enter the tube and make it expand without an important increase of pressure.
The total volume of the tube is $V= \pi r^2 L$ and the volume of the part with sugar of concentration $c_0$ is $V= \pi r^2 x$. In the simplest case, we assume that the flow inward only up in he sugar interval $[0,x]$. The rest of the tube just expands without any restoring force ($E=0$)

\begin{equation}
{\frac{d V}{d t}}=2 \pi L r {\frac{d r}{d t}}=  \pi r^2 {\frac{d x}{d t}} + 2 \pi r x {\frac{d r}{d t}}
\end{equation}
Thus
\begin{equation}
\label{x}
(L-x)r^2 = ( L-x_0) r_0^2
\end{equation}
The inflow is
\begin{equation}
I = {\frac{d V}{d t}} = 2 \pi r  \int_0^xL_W R T c(x') d x'= 2 \pi r L_w R T c x
\end{equation}
But $c$ changes so that $c x r^2 = c_0 x_0 r_0^2$ and thus
\begin{equation}
 {\frac{d V}{d t}} = 2 \pi L r {\frac{d r}{d t}}= {\frac{2 \pi L_w R T c_0 x_0 r_0^2}{r}}
\end{equation}
so
\begin{equation}
 {\frac{d r}{d t}}= {\frac{ L_w R T c_0 x_0 r_0^2}{L r^2}}
\end{equation}
Note that $c_0$ os not the average over the whole tube. This would be $\bar{c_0} = {\frac{x_0}{L}}c_0$.
We have earlier introduced the time scale
\begin{equation}
\tau= {\frac{ r_0}{2 L_w R T c_0 }}
\end{equation}
Using instead
\begin{equation}
\bar{\tau}= {\frac{ r_0}{2 L_w R T \bar{c_0} }}
\end{equation}
letting $t= s \bar{\tau}$ and $r=y r_0$ we get
\begin{equation}
y'(s)= {\frac{ 1}{2 y^2}}
\end{equation}
with solution
\begin{equation}
y(s)= \left({\frac{ 3 s}{2 }}  + 1\right)^{1/3}
\end{equation}

Now $x$ can be found from (\ref{x}):
\begin{equation}
(L-x)y^2 = ( L-x_0)
\end{equation}
so
\begin{equation}
x= L- ( L-x_0)  \left({\frac{ 3 s}{2 }} + 1\right)^{-2/3} =  L- ( L-x_0)  \left({\frac{ 3 t}{2 \tau }}  + 1\right)^{-2/3}
\end{equation}
Now the  front velocity is
\begin{equation}
v(t)= x'(t)= {\frac{ L-x_0}{\tau}} \left({\frac{ 3 t}{2 \tau }}  + 1\right)^{-t/3}
\end{equation}
and initially it takes the value
\begin{equation}
v_0= x'(0)= {\frac{ L-x_0}{\tau}}
\end{equation}

We now assume that the influx of water creates an increased pressure, and that, consequently water can flow out of the sugarless region.
We assume that the elastic properties are governed by
\begin{equation}
p(r)=d_0 E  {\frac{ r -r_0}{r^2}}
\end{equation}
where $d_0$ is the initial width of the membrane, $E$ is Young's modulus and we have assumed incompressibility (i.e. Poisson's ratio $\nu = 1/2$) and, so $d r = d_0 r_0$. Now the equation for the current is
\begin{equation}
I = {\frac{d V}{d t}} = 2 \pi r L_w  \int_0^L \left(  R T c(x') -p(x') \right) d x'= 2 \pi r L_w (R T c \, x-p L)
\end{equation}
where we have assumed (as in our experiments) that the pressure is constant along the tube. Thus
\begin{equation}
 {\frac{d V}{d t}} = 2 \pi r L_W \left(R T c \, x- d_0 E  {\frac{L( r -r_0)}{r^2}} \right)
\end{equation}
Again,  $c x r^2 = c_0 x_0 r_0^2$ so
\begin{equation}
 {\frac{d V}{d t}} = 2 \pi L r {\frac{d r}{d t}}=  2 \pi r L_w \left( {\frac{ R T c_0 x_0 r_0^2}{r^2}} - d_0 E {\frac{L ( r -r_0)}{ r^2}} \right)
\end{equation}
so
\begin{equation}
 {\frac{d r}{d t}}= {\frac{ L_w R T \bar{c_0}  r_0^2}{ r^2}}-  {\frac{L_w d_0 E ( r -r_0)}{ r^2}}
\end{equation}
Using again $r = r_0 y$ and $t= \bar{\tau} s$ we get the dimensionless equation
\begin{equation}
\label{elas}
 {\frac{d y}{d s}}= {\frac{1}{ 2 }}\left( {\frac{1- B\, (y-1)}{  y^2}} \right)
\end{equation}
where
\begin{equation}
B={\frac{2  \bar{\tau} d_0 E L_w}{  r_0^2}} =  {\frac{d_0}{ r_0}} {\frac{E}{ R T \bar{c_0}}}
\end{equation}
In this case we see that the radius saturates for long times to the value:
\begin{equation}
r_f = r_0 y_f = r_0 {\frac{1+B}{B}}
\end{equation}
and $B$ can be expressed as
\begin{equation}
B= {\frac{1}{y_f -1}}
\end{equation}
This equation can be solved for $s(y)$ by inverting:
\begin{equation}
 {\frac{d s}{d y}}= 2{\frac{y^2}{ 1+B-By}}
\end{equation}
which can be integrated (with the initial condition $s(1)=0$) to
\begin{equation}
\label{inv}
s(y) = {\frac{1}{B^3}}\left(-(B (y-1) (2 + B (3 + y)) + 2 (1 + B)^2 \log (1 + B - B y) \right)
\end{equation}
valid as long as $y<y_f$. For small $B$ this is is approximately:
\begin{equation}
s(y) = 1/3 (-1 + y^3) + 1/12 (1 - 4 y^3 + 3 y^4) B + o(B^2)
\end{equation}
and thus agrees with the result of last section, i. e. the result for $E \to 0$. For larger $B$ the part of the integral close to

To find the front velocity $v_f = \dot{x}$ we look at the volume $V_2 = \pi r^2 (L-x)$ above the sugar front.
By definition, the sugar concentration is zero there, so the water flux through the membrane is only due to the pressure difference:
\begin{equation}
 {\frac{d V_2}{d t}} =2 \pi r (L-x) {\frac{d r}{d t}} -\pi r^2 {\frac{d x}{d t}}=-p(L-x)L_w 2 \pi r
\end{equation}
so
\begin{equation}
{\frac{1}{L-x}} {\frac{d x}{d t}}= - {\frac{d \log (L-x)}{d t}}
={\frac{2} {r}} \left({\frac{d r}{d t}} +L_w 2  d_0 E  {\frac{( r -r_0)}{r^2}} \right)
\end{equation}
or
\begin{eqnarray}
\nonumber
 {\frac{d \log (L-x)}{d s}}
&=&-{\frac{2} {y}} \left({\frac{d y}{d s}} + \tau r_0^{-2} L_w 2  d_0 E  {\frac{( y -1)}{y^2}} \right)\\
&=&-{\frac{2} {y}} \left({\frac{d y}{d s}} + B {\frac{( y -1)}{y^2}} \right)
\end{eqnarray}
from which we can find $x(t)$ from the solution for $r(t)$, although we cannot write it explicitly since the explicit form of $r(t)$ is not known (only the inverse (\ref{inv})). For $B=0$ we again recover the flappy limit (E=0), where $L-x \sim y^{-2}$.

\bibliographystyle{jfm}	
\bibliography{Osmomain}		
\end{document}